\documentclass[preprint,double-spaced,floatfix ,showpacs,preprintnumbers,
amsmath,amssymb,prd]{revtex4}

\usepackage{graphicx}
\usepackage{dcolumn}
\usepackage{bm}
\usepackage{color}

\newcommand{\eref}[1]{Eqs.~(\ref{#1})}
\newcommand{\sref}[1]{section~\ref{#1}}
\newcommand{\cref}[1]{chapter~\ref{#1}}
\newcommand{\fref}[1]{figure~\ref{#1}}
\newcommand{\tref}[1]{table~\ref{#1}}
\newcommand{\Eref}[1]{Equation (\ref{#1})}

\newcommand{\Cref}[1]{Chapter~\ref{#1}}
\newcommand{\Fref}[1]{Figure~\ref{#1}}

\newcommand{\beq}[1] {\begin{equation} \label{#1}}

\newcommand{\eeq} {\end{equation}}

\begin{document}

\title{Complete model of a spherical gravitational wave detector with
 capacitive transducers. Calibration and sensitivity optimization.}

\author{Luciano Gottardi}
 \affiliation{LION, Institute of Physics, Kamerlingh Onnes Laboratorium,
 Leiden University, Leiden, The Netherlands}
 \altaffiliation[Current address: ]{SRON, National Institute for Space Research, High Energy Astrophysics Division, Utrecht, the Netherlands}

\email{l.gottardi@sron.nl}

\date{\today}

\begin{abstract}

We  report the  results of  a detailed  numerical analysis  of  a real
resonant  spherical  gravitational  wave  antenna operating  with  six
resonant two-mode  capacitive transducers read  out by superconducting
quantum interference  devices (SQUID) amplifiers.  We derive  a set of
equations   to  describe  the   electro-mechanical  dynamics   of  the
detector. The  model takes  into account the  effect of all  the noise
sources present in each transducer chain: the thermal noise associated
with   the  mechanical   resonators,  the   thermal  noise   from  the
superconducting impedance matching  transformer, the back-action noise
and the additive current noise of the SQUID amplifier.  Asymmetries in
the  detector   signal-to-noise  ratio  and   bandwidth,  coming  from
considering the  transducers not as  point-like objects but  as sensor
with physically defined geometry and dimension, are also investigated.
We calculate the sensitivity for an ultracryogenic, 30 ton, 2 meter in
diameter,  spherical detector with  optimal and  non-optimal impedance
matching  of   the  electrical  read-out  scheme   to  the  mechanical
modes.  The results of  the analysis  is useful  not only  to optimize
existing smaller  mass spherical  detector like MiniGrail,  in Leiden,
but also  as a technological  guideline for future  massive detectors.
Furthermore we calculate the antenna patterns when the sphere operates
with one, three and six transducers. The sky coverage for two detectors
based in  The Netherlands and  Brasil and operating in  coincidence is
also  estimated.   Finally,  we  describe  and  numerically  verify  a
calibration  and   filtering  procedure  useful   for  diagnostic  and
detection purposes  in analogy  with existing resonant  bar detectors.

\end{abstract}

\maketitle

\section{Introduction}

 Resonant bar  antennas are  in continuous operation  at $4.2  K$ with
sensitivity      and      bandwidth      never     reached      before
\cite{Baggio05},\cite{EXPNAU}.   These detectors  could  improve their
sensitivity  of  one order  of  magnitude  in  the coming  years  when
operating   at   $100  mK$.    The   first  ultracryogenic   spherical
gravitational  wave  detectors  \cite{MINIGRAIL,SCHENBERG}  are
currently  completing  their  engineering   phase  and  will  soon  be
operational  with  an   expected  sensitivity  better  than  $10^{-21}
Hz^{-1/2}$ at $3 kHz$.  Much interest is now directed towards the next
generation  of acoustic detectors,  which will  be large  mass spheres
\cite{SFERA}  equipped  with   traditional  resonant  transducers  and
broadband  dual  resonators  \cite{DUAL,DUAL06}.  The  resonant
spheres rely on available technology \cite{Baggio05,EXPNAU}, a
rather              extensive             theoretical             work
\cite{AshDrei75,WagPaik77,Merkowitz98,Lobo95,ZhouMich95,Stevenson98},     
and    the
ground-breaking   experimental   work   performed   so  far   on   the
ultracryogenic  small  mass  sphere  Minigrail  \cite{MINIGRAIL}.  The
wide-band  dual detector  potentially solves  the problem  of relative
narrow bandwidth of current resonant bar and spheres, but the required
technology needs to be assessed in separate experiments \cite{DUAL06}.
Here  we numerically  analyse the  sensitivity and  performance  of an
ultracryogenic   spherical   detector,   equipped   with   capacitive,
SQUID-based, resonant transducers.  A  general analysis of the problem
of  the   read-out  system  of   for  linear  detector  is   given  in
\cite{Price87,PallPizz81}.                Previous               works
\cite{Merkowitz95,LoboSerr96,Lobo00,ZhouMich95,Stevenson98,Merkowitz98} 
have provided a solution for the
read-out  and  inverse  problem  of  a  spherical  detector,  deriving
equations  of motion  for  the five  degenerate quadrupole  spheroidal
modes  coupled  to   N  identical,  point-like,  single-mode  resonant
transducers  located at arbitrary  points on  the sphere  surface. The
effects  of  transducer  asymmetries  on the  strain  sensitivity  and
bandwidth  was  studied  in \cite{Merkowitz98,Stevenson98}  for
rather   generic   radial,    point-like,   single   mode,   identical
resonators. In  \cite{Harry01} the strain sensitivity  for a spherical
gravitational  wave detector  with a  three-mode  inductive transducer
with optimal parameters is calculated.

 The  present  paper  shows   the  results  of  a  detailed  numerical
calculation  of the performance  of a  spherical detector,  which uses
2-mode capacitive transducers where  the electrical resonant mode of a
superconducting  matching  network  can   be  tuned  to  the  resonant
mechanical  modes.   The   signal  current  from  the  superconducting
matching transformer  is read-out  by sensitive SQUID  amplifiers.  We
chose  this transduction system  mainly for  two reasons:  first, most
existing bar and spherical antenna use capacitive transducers, second,
the  technology involved  is so  far the  most advanced.   As recently
experimentally demonstrated  on the currently  most sensitive resonant
antenna AURIGA  \cite{Baggio05}, such a read-out  scheme enhances both
the sensitivity and the bandwidth  of a resonant detector when working
in  the tuned  mode.  Two-stage  SQUID amplifiers  coupled  to a  high
quality factor  load can reach  nowadays an energy resolution  only an
order of magnitude higher than  its quantum limit when properly cooled
down to $100 mK$ \cite{Falferi06}. Two-stage SQUID amplifiers operates
at 4-5K on the bar antenna AURIGA \cite{Baggio05}, and on the spherical
antenna  MiniGRAIL \cite{GottardiPhD},  with  an unprecedently  reached
sensitivity of about $600 \hbar$.
  
  We use the analysis presented here to study the detector sensitivity
as a function of the  SQUIDs, the superconducting matching network and
the  mechanical resonators  intrinsic  parameters, and  to define  the
optimal  coupling  between antenna,  transducer  and amplifiers.   The
simulations consider  the effect of  all the parameters involved  in a
real detector, including  the effect of the cold  damping network used
in the fluxed lock-loop (FLL)  of the SQUID amplifier to stabilize the
read-out  of   high  Q  loads\cite{Vinante02}.   We   study  also  the
sensitivity  and  signal   bandwidth  deterioration  coming  from  the
transducer  being  a  geometrically  extended  object  rather  than  a
point-like mass.  Finally we describe and numerically test a method to
fully calibrate a spherical detector and to derived the optimal filter
parameters from the experimental data. This method is a generalization
of the one used with the resonant bar antenna AURIGA\cite{Baggio02}.
\noindent The codes generated to  perform such an analysis can be used
as a  guideline for the development  of future detectors as  well as a
tool  to evaluate  the  performance of  present  small mass  spherical
detectors.

 In  \sref{sphereGW_equation} we  give  an overview  of the  equations
necessary to describe the coupling of a gravitational wave to a sphere
and  of a  sphere to  N mechanical  resonator following  the formalism
introduced  by Johnson and  Merkowitz \cite{Merkowitz95}.  We complete
the equation of motion for  a capacitive transducer coupled to a SQUID
amplifier  through   a  superconducting  matching   transformer.   The
detector  strain sensitivity, noise  temperature and  signal bandwidth
are  derived  using  the   generilized  vector  approach  proposed  by
Stevenson  \cite{Stevenson98}. This  method  is particularly  powerful
when  transducers  are not  identical  not  only  in their  mechanical
parameters, but also with respect to their noise sources. It indicates
a rather  simple method to  handle the correlation  between transducer
output channels  and to  form statistically independent  channels (the
{\it mode  channel} concept used in \cite{Merkowitz95})  in absence of
symmetries.  In \sref{noisesources} we  describe all the noise sources
acting on the sphere and the transducer chain.
\noindent  The  results of  the  numerical  analysis  are reported  in
\sref{analysis}.  First,  following  the  description of  a  parameter
optimization  procedure, we  calculate the  strain sensitivity  for an
ultracryogenic spherical detector,  2 m in diameter, made  of a copper
and  alluminum  alloy  ($CuAl6\%$),  and  equipped with  one  and  six
transducers located in the positions of the Truncated Icosahedral (TI)
configuration      proposed     by      Johnson      and     Merkowitz
\cite{Merkowitz95}. Secondly, the  antenna patterns are calculated for
two spherical detectors with  optimal sensitivity located in the north
and south hemisphere  of the earth. The sky  coverage is estimated for
such spheres operating  with one, three and six  resonators, both when
working independently  and in coincidence. Finally, the  effect on the
sensitivity is evaluated  when non-identical and non-ideal transducers
are employed.  In  \sref{calibration} a complete calibration procedure
for a spherical detector is described.  First we show a way to derived
experimentally the  equivalent temperature of the  mechanical modes
of a sphere equipped with six transducers. Then we discuss a procedure
to measure the detector transfer  functions and to calibrate its force
response. Such a method is  finally tested by calculating the detector
{\it mode channels} response with simulated gravitational wave bursts.
Finally, \sref{conclusions} summarizes the results.
\section{Spherical GW resonant detectors}
\label{sphereGW_equation} In  this section we describe  the model used
to derive  the sensitivity of  a spherical gravitational  wave antenna
with  N transducers  coupled  only  to radial  motion.  We consider  a
capacitive transducer, which can be operated in a 2-mode configuration
when the electrical mode is tuned to the mechanical one.

The dynamics of a sphere coupled to radial transducers and interacting
with  gravitational  waves is  described  below  using  the {\it  mode
channels}    formalism   introduced    by   Johnson    and   Merkowitz
\cite{Merkowitz95},  and  matrices  notation  suitable  to  derive  the
signal-to-noise   ratio,  SNR,   and  the   noise  temperature   of  a
multichannel system \cite{Stevenson98}.
\subsection{Coupling of a bare sphere with the gravitational field}

A  gravitational wave  is  a time-dependent  deviation  of the  metric
perturbation. In  the coordinate  frame of the  wave, denoted  here by
primed coordinates and indices, with the origin at the detector center
of mass and  the {\it z'} axis aligned  with the propagation direction
of the wave, the spatial metric  perturbation in the TT gauge is given
by   
\beq{GWH}   
\mathbf{H'}(t)=\left({\begin{array}{ccc}   h_+(t)   &
h_\times (t) & 0 \\ \\ h_\times(t) & -h_+ (t) & 0 \\ \\ 0& 0 & 0 \\ \\
\end{array}}\right), \eeq  with $h_+$ and  $h_\times$ corresponding to
the two independent wave polarizations.  A  wave with polarization
$h_+$ deforms a test ring into an ellipse with axes in the $x$ and $y$
directions. A wave with polarization  $h_\times$ deforms the ring at a
45-degree angle to the $x$  and $y$ directions. A circularly polarized
wave has  $h_+ = \pm h_\times$  and rotates the deformation  of a test
ring    in     the    right-handed    (or     left-handed)    direction
\cite{MISTHORWHE73}.

The dynamics  of a bare resonant  sphere can be  described by ordinary
elastic  theory.  A  forces  $f$  acting  on the  body  will  cause  a
displacement   of  the   sphere  mass   element  at   its  equilibrium
position.  The mechanics  is described  by the  standard  equations of
motion of a forced oscillator.   In this section we limit ourselves to
review  the main  result. More  complete  treatments can  be found  in
\cite{AshDrei75,WagPaik77,Merkowitz95,ZhouMich95,Lobo95}.            In
particular  we  make use  of  the  formalism  introduced by  Merkowitz
\cite{Merkowitz95,Merkowitz98}.

Denoting by ${\bf f}({\bf x},t)$ the total force that acts on the sphere,
including the gravitational wave force,  at
the position ${\bf x}=\{x,y,z\}$ and time $t$, one finds that the 
equation for each mode
amplitude is the one for a forced harmonic oscillator. After
Fourier transforming,  each mode amplitude can be written as
\begin{equation}
\label{modeamplFT}
\begin{array}{c}
  {\bf a_m}(\omega)= \frac{3}{4 \rho
  \pi R^3}\frac{1}{\omega_m^2-\omega^2
  +j\omega_m^2\Phi_m}\times \\
\times \int {\mathbf{\Psi}_{lm}(\mathbf{x}) 
    \mathbf{f} \left( {\mathbf{x,\omega}} \right)d^3x},
\end{array} 
\end{equation}
where R is the radius of the sphere and $\rho$ the
density. On the right hand side of the equation, the first
factor is an arbitrary normalization constant.
 The second factor describes the oscillating
 nature of  the displacement where $\omega_m$ and $\Phi_m=1/Q_m$  are,
 respectively the  resonance frequency and the loss angle associated with the
 quality factor  $Q_m$ of the $m$th mode. The
 integral is calculated over the entire volume of the sphere where
$\mathbf{\Psi}_{l,m}\left( {\mathbf{x}} \right)$ are the time independent
 orthogonal elastic eigenfunctions of the sphere with $l=0$ or $l=2$ .

In general relativity  only 5 quadrupolar  modes  of 
vibration ($\ell=2$) will strongly couple to the force density of a 
gravitational wave due to the fact that the tensor $\mathbf{H'}$ is traceless.
 In a perfect sphere they are all degenerate, having the same angular 
eigenfrequency $\omega_0$. 
The quadrupolar modes can be written in terms of
the convenient set of the  five real spherical harmonics $Y_{m}(\theta,\phi)$, 
which are defined as follows:
\beq{sph_harm}
\begin{array}{r}
\left(\begin{array}{l}
Y_{1}\\
Y_{2}\\
Y_{3}\\
Y_{4}\\
Y_{5}\\
\end{array}\right)
=\sqrt{\frac{15}{16\pi}}
\left(\begin{array}{c}
cos 2\phi sin^2\theta\\
sin2\phi sin^2\theta\\
sin\phi sin2\theta\\
cos\phi sin2\theta\\
\frac{1}{\sqrt3}(3cos^2\theta -1).\\
\end{array}\right)
\end{array}
\eeq

They are the result of a linear combination of the usual complex-valued
 spherical harmonics $Y_{2m}$.

 For a sphere of radius $R$ the 
eigenfunctions can be written as: 
\begin{equation}
  \mathbf{\Psi}_{m} = \left[ {\alpha (r) 
  \mathbf{\hat r} + \beta(r) R \mathbf{\nabla} } \right]
  Y_{m}(\theta, \phi).   
   \label{wavefunction}
\end{equation}
The radial eigenfunctions $\alpha \left( r \right)$ and $\beta  \left( r \right)$ 
determine the motion in the radial and tangential directions, respectively.  
An explicit description of the motion in the radial and tangential directions is
given by Ashby and Dreitlein \cite{AshDrei75}.

In  the lab-frame with origin at the center of mass of the detector and the z
 axis aligned with the local vertical, a  gravitational wave produces an
 effective time dependent {\it tidal}  
force $\mathbf{F_{m}^S}$ on each mode $m$ of the sphere equal to the overlap
  integral of \eref{modeamplFT}. 
One finds
\begin{eqnarray}
   F_m^S(t) 
   & = & 
   \sqrt{\frac{4\pi}{15}} \rho \ddot{h}_m(t) R^4 \left[ 
   {c \, J_2\left( {qR} \right)+3d \, J_2\left( {kR} \right)} 
   \right] \nonumber \\
   & = & 
   \frac{1}{2} \ddot{h}_m(t) \, m_S \, \chi R,
\label{effforce}
\end{eqnarray}
\noindent  where  $J_2$  is  the spherical  Bessel  function  of  order  2,  
the
coefficients $c, \, d$ specify the shape of the eigenfunctions and are
weakly dependent on  the material Poisson ratio \cite{WagPaik77}, $m_S$
is the  physical mass  of the sphere  and $R  \chi $ is  the effective
length of each  mode where $\chi$ depends on the  Poisson ratio and is
equal  to  0.327 for  the  CuAl  sphere  considered in  the  following
analysis. $h_m$ are the {\it spherical amplitudes} \cite{Merkowitz95},
a   complete    and   orthogonal   representation    of   the   metric
perturbation.  $q$  and  $k$  are respectively  the  longitudinal  and
transverse wave  vectors as defined in  \cite{AshDrei75}. The effective
force  $F^S_m$ on  the corresponding  mode  of a  sphere is  therefore
uniquely determined  by each spherical component  of the gravitational
field.

The force acting on each spheroidal  mode $m$ in the lab frame can now
be   written   in  terms   of   the   gravitational  wave   amplitudes
\beq{Fmlabframe}  \mathbf{F_m^S}=\frac{1}{2}m_S  \chi  R  \omega^2  \;
\mathbf{T_V} \left(\begin{array}{c}  h_+\\ h_\times \end{array}\right), 
\eeq
where $\mathbf{T_V}$, given by 
\beq{TV}
\begin{array}{c}         \mathbf{T_V}=\left({        \begin{array}{cc}
\frac{1}{2}(1+cos^2\theta)cos  2\phi   &  cos  \theta   sin  2\phi  \\
-\frac{1}{2}(1+cos^2\theta)sin  2\phi  &   cos  \theta  cos  2\phi  \\
-\frac{1}{2}sin   2\theta   sin  \phi   &   sin\theta   cos  \phi   \\
\frac{1}{2}sin   2\theta    cos   \phi   &    sin\theta   sin\phi   \\
\frac{\sqrt{3}}{2} sin^2 \theta & 0
         \end{array} }\right)\times \\ \\ \times\left({
\begin{array}{cc} cos 2\psi & sin 2\psi \\ -sin 2\psi & cos 2\psi
\end{array} }\right),
\end{array}  \eeq is  the  transformation matrix,  which converts  the
gravitational  wave  amplitude  in   the  wave  frame  into  spherical
amplitude  in the lab  frame.  Here  we used  the y-convention  of the
Euler angles  shown in \fref{sphere_TI} and the  linear combination of
the  spherical  harmonics  described  in \eref{sph_harm}.   The  angle
$\psi$ is  the first Euler angle  in the rotation  relating the wave
frame to the laboratory frame  and it carries information about the GW
polarization.

\subsection{Sphere with N resonant transducers}

Resonant transducers  are used on  resonant detectors, either  bars or
spheres \cite{Baggio05,Astone03,MINIGRAIL}, in  order to improve their
sensitivity  and  bandwidth.   We  consider  here  the  same  type  of
transducers  as displacement  sensors.  They  consist of  a mechanical
resonator with the fundamental mode  tuned to the quadrupolar modes of
the antenna. At resonance, there is a transfer of momentum between the
resonator  and the  antenna, turning  small displacements  of  a large
antenna into large displacements of the small resonator.

Let us  consider a  set of  $N$ resonators attached  to the  sphere at
arbitrary positions  $(\theta_j, \Phi_j)$. The values  of the relative
radial displacement of  the sphere at the transducers  location can be
grouped  together   into  {\it  pattern  vectors}   for  a  particular
mode. These  column vectors, in  turn, may be  put together to  form a
{\it pattern matrix}  $B_{mj}$ defined by \cite{Merkowitz97}, 
\beq{Bmj}
B_{mj}=\frac{1}{\alpha}\mathbf{r}\cdot
\mathbf{\Psi}_m(\theta_j,\phi_j), 
\eeq
\noindent  where $\alpha$  is the  radial eigenfunction  introduced in
\eref{wavefunction}. One gets 
\beq{BmjYm} 
B_{mj}=Y_m(\theta_j,\phi_j).
\eeq  
Assuming  that  each  resonator  is designed  to  obey  the  one
dimensional harmonic  oscillator law, the coupled  equations of motion
for the sphere modes, written in matrix form, are
\begin{equation}
\begin{array}{c}
   \left[{ \begin{array}{cc}
      m_S\mathbf{I} & 
      \mathbf{0} \\
      m_R\alpha\mathbf{ B}^T & 
      m_R\mathbf{ I}
   \end{array} }\right]
   \left[ \begin{array}{c}
      \mathbf{\ddot a}(t) \\
      \mathbf{\ddot q}(t)
   \end{array} \right] +
   \left[ \begin{array}{cc}
      k_S\mathbf{ I} & 
      -k_R\alpha\mathbf{ B}\\
      \mathbf{ 0} & 
      k_R\mathbf{ I}
   \end{array} \right]
   \left[ \begin{array}{c}
      \mathbf{ a}(t) \\
      \mathbf{ q}(t)
   \end{array} \right] =  \\ \\
   = \left[ \begin{array}{cc}
      \mathbf{ I} & 
      -\alpha \mathbf{ B} \\
      \mathbf{ 0} & 
      \mathbf{ I}
   \end{array} \right]
   \left[ \begin{array}{c}
      \mathbf{F^S}(t) \\
      \mathbf{F^R}(t)
   \end{array} \right],
   \end{array}
\label{eqn_of_motion}
\end{equation} where  matrices are denoted  by bold fonts  and capital
letters and  vectors by bold fonts  and low case  letters.  The vector
$\mathbf{a}$  has 5  components and  the vector  $\mathbf{q}$  has $N$
components. They  represent the radial displacement of  the sphere and
the  resonator,  respectively. For identical  sphere  modes  and
identical  transducers, the factors  $m$, $k$,  indicating respectively
mass  and  spring  constants  are  identical and  can  be  treated  as
numbers. In  reality, each mode has  its own mass,  quality factor and
spring  constant, so  they have  the form  of a  diagonal  matrix with
components  $m_i^s$,  $k_i^s=m_i^s \omega_i^2(1+i\Phi_i(\omega))$  for
the   sphere,    with   $i=1..5$   and    $m_j^r$   and   $k_j^r=m_j^r
\omega_j^2(1+i\Phi_j(\omega)))$  for  the  resonators, with  $j=1..N$.
Here  $\omega_n$,   with  $n=i,j$,   is  the  natural   frequency  and
$\Phi_n(\omega)$  is  the  loss  angle  of  each  resonant  mode.   It
represents  the frequency  dependence  of  the loss  of  a mode.   For
commonly observed dissipations  in metals \cite{Yamamoto02}, losses do
not depends on frequency and $\Phi_n(\omega)=1/Q_n$ where $Q_n$ is the
mode quality factor.  In the  case of viscous damping, due for example
to  eddy-current  effects,  the  loss  angle is  proportional  to  the
frequency  and is  given by  $\Phi_n(\omega)=\omega/\omega_n  Q_n$. We
consider the first dissipation mechanism to describe the losses in the
mechanical modes.   $\mathbf{F^S}$ and $\mathbf{F^R}$  are the driving
forces, which include the  gravitational waves contribution as well as
the forces generated by noise sources.

\begin{figure}
 \includegraphics[scale=0.75]{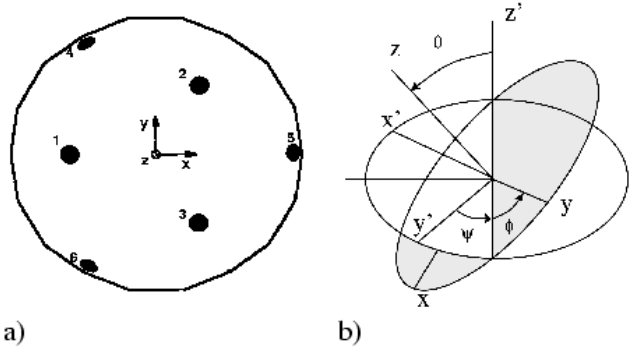}
 \caption{\label{sphere_TI}  $\mathbf{a)}$  The truncated  icosahedral
(TI)  arrangement  for a  spherical  gravitational  wave antenna  with
resonator  locations  indicated.   The  numbering  of  the  resonators
corresponds  to  the  ordering  used in  the  numerical  calculations.
$\mathbf{b)}$ Euler angle transforms convention.}
\end{figure}

\Eref{eqn_of_motion}    fully   describes   the    mechanical   system
sphere-resonators when the {\it pattern matrix} $B_{mj}$ is known. 
Here we consider the special transducer configuration proposed by Johnson and
Merkowitz  \cite{JohMerk93,Merkowitz95}. It  consists  of a  set of
six  transducers placed  on  the  6 pentagonal  faces  of a  Truncated
Icosahedron  (TI). The  resonators are  located at  two  polar angles,
$\theta_{TI}   =37.3773^{\circ}$   and   $79.1876^{\circ}$   as illustrated   
in \fref{sphere_TI}. Their azimuthal  angles $\psi_{TI}$ are multiples of
$60^{\circ}$.

\begin{figure}
 \includegraphics[scale=0.7]{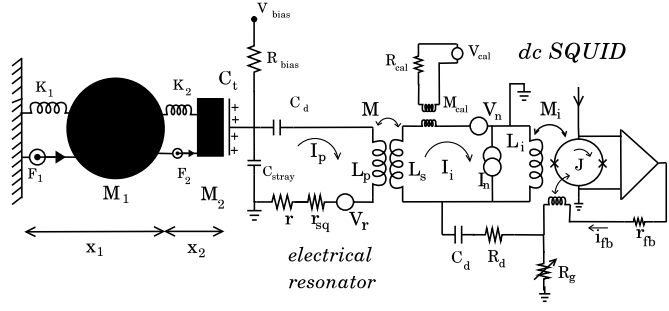}
 \caption{\label{sphere_twomode_cap}  Single  mode  electro-mechanical
model of a spherical  antenna with mechanical resonator and capacitive
transducer  coupled  to a  SQUID  through  a superconducting  matching
transformer.}
\end{figure}

 Below we
derive  the complete  equations  of motion  for  a spherical  detector
equipped with capacitive transducer and SQUID amplifiers. 
 In  a  capacitive
transducer,  the  resonating  mass,  tuned to  the  spheroidal  modes,
modulates the charge  of a parallel plate capacitor  biased at a large
constant voltage. The  capacitor is formed by the  resonating mass top
surface and an electrode, assembled with a gap of the order of tens of
micrometers.  The  input coil of the  dc-SQUID chip is  coupled to the
capacitive transducer via  a high-{\it Q} superconductive transformer,
which  can have,  eventually, the  electric resonance  coupled  to the
mechanical   modes   in  order   to   enhance   the  bandwidth.    The
superconducting transformer is essential to match the low impedance of
the  SQUID with the  high impedance  of the  capacitor. In  the model,
which is schematically  shown in \fref{sphere_twomode_cap}, we include
the relevant Gaussian noise sources of the read-out scheme.

\subsection{Equation of  motion of a spherical  detector with resonant
capacitive transducers}

The motion  equations of  a single
capacitive transducer coupled to one mechanical mode of the sphere can
be generalized to include the complete mechanical response of a sphere
coupled to $N$ resonators,  described by \eref{eqn_of_motion}, and all
the  equations for  the electrical  circuit of  each  resonator. After
Fourier transforming, \eref{eqn_of_motion}  can be simplified defining
the following $5+N$ square matrices: \beq{matricesMKH}
\begin{array}{l} \mathbf{M}=\left[{  \begin{array}{cc} m_S\mathbf{I} &
\mathbf{0} \\ m_R\alpha\mathbf{ B}^T & m_R\mathbf{ I}
   \end{array}  }\right],  \quad  \mathbf{K}=\left[  \begin{array}{cc}
k_S\mathbf{ I} & -k_R\alpha\mathbf{ B}\\ \mathbf{ 0} & k_R\mathbf{ I}
   \end{array}  \right],\\  \\  \mathbf{A}=  \left[  \begin{array}{cc}
\mathbf{ I} & -\alpha \mathbf{ B} \\ \mathbf{ 0} & \mathbf{ I}
   \end{array} \right].
   \end{array} \eeq

They  are respectively  the mass,  elastic and  force matrices  of the
coupled sphere. The force matrix $\mathbf{A}$ describes the mechanical
coupling  between the  5+N  resonant  modes of  the  detector. We  can
finally                     write                     \beq{Mechmatrix}
\left[-\omega^2\mathbf{M}+\mathbf{K}\right]\left[      \begin{array}{c}
\mathbf{a}(\omega)\\ \mathbf{q}(\omega)
   \end{array}        \right]=\mathbf{A}\left[        \begin{array}{c}
\mathbf{F_N^S}(\omega) \\ \mathbf{F_N^R}(\omega)
   \end{array}  \right].   \eeq  We   assume  that  each  of  the  $N$
transducers  mounted on  the sphere  has the  same  electrical circuit
configuration   described   in   \fref{sphere_twomode_cap},  but   not
necessarily the same value of the parameters. We shall define then the
vectors                      $\mathbf{I_p(\omega)}=(I_{p,1}..I_{p,N})$,
$\mathbf{I_i(\omega)}=(I_{i,1}                             ..I_{i,N})$,
$\mathbf{V_r(\omega)}=(V_{r,1}..V_{r,N})$,
$\mathbf{V_n(\omega)}=(V_{r,1}..V_{n,N})$,  which  describe, for  each
transducer, respectively  the current in  the superconducting matching
transformer, the current in the  input coil of the dc SQUID amplifier,
the voltage noise generated in the $LC$ superconducting resonators and
the voltage noise of the dc SQUID amplifiers.

Denoting by  $E_i$ the electric  field stored in the  $i$th capacitive
transducer,  we   define  the   electric  field  matrix   as  follows:
\beq{Ematrix} 
\mathbf{E}=diag(E_1..E_N).  
\eeq  
The electrical circuit equations can be written in matrix form as
follows 
\beq{El_eqn_of_motion} \left[{ \begin{array}{ccc} \mathbf{E} &
\mathbf{Z_{11}}  &  \mathbf{Z_{12}}\\  
\mathbf{0}  &\mathbf{Z_{21}}  &
\mathbf{Z_{22}}
   \end{array}  }\right]\left[ \begin{array}{c}  \mathbf{q}(\omega) \\
\mathbf{I_{p}}(\omega)                                               \\
\mathbf{I_i}(\omega)\end{array}\right]=\left[         \begin{array}{cc}
\mathbf{ I} & \mathbf{0} \\ 
\mathbf{ 0} & \mathbf{ I}
   \end{array} \right]  \left[ \begin{array}{c} \mathbf{V_{r}}(\omega)
\\ 
\mathbf{V_n}(\omega)
   \end{array} \right].  
\eeq

We chose the  electric matrix $\mathbf{E}$ to be  diagonal, because we
made the reasonable assumption that the electric field force acts only
on the resonator which the  field is applied to.  The impedance matrix
\beq{Zmatrix}  \mathbf{Z}=\left[{ \begin{array}{cc}  \mathbf{Z_{11}} &
\mathbf{Z_{12}}\\ \mathbf{Z_{21}} & \mathbf{Z_{22}}
   \end{array} }\right] \eeq is  a $(N+N)\times (N+N)$ matrix. Each of
the four  $N \times  N$ matrices $\mathbf{Z_{ij}}$  is diagonal  if we
consider  as negligible  the possible  crosstalk between  the read-out
electronics   of   each    transducer.   Each   diagonal   member   of
$\mathbf{Z_{ij}}$ is equal to
\beq{Zij_members}
\begin{array}{l}  {Z^i_{11}}=r^i+r_s^i+j\omega  L_p^i+\frac{1}{j\omega
C_{p}^i}, \qquad  {Z^i_{12}}= -j\omega M^i \\  \\ {Z^i_{21}}= -j\omega
M^i , \qquad {Z^i_{22}}=j\omega\left(L_s^i+L_{in}^i \right),
\end{array}  \eeq where  $r^i$  is a  resistance  associated with  the
losses  in  the  superconducting  resonator. $C_{p}^i$  is  the  total
transducer  capacitance  resulting   from  the  parallel  between  the
transducer and  parasitic capacitance. $r_s$ is  a lossless resistance
resulting  from operating  the  SQUID amplifier  in  flux locked  loop
(FLL).  There is  no thermal  noise contribution  associated  to $r_s$
because it  is the result of  a feedback mechanism.  Such a resistance
can   be   controlled    by   implementing   a   cold-damping   system
\cite{Vinante02}. In this  way the detector has a  virtual low quality
factor making the FLL  electrically stable. We remark that introducing
such  a damping  scheme in  our  calculations brought  benefit to  the
numerical analysis, by eliminating  the computational problem of sharp
resonances.

To  fully  describe the  mechanical  and  electrical  dynamics of  the
detector  we  have to  introduce  the  back-action  of the  electrical
read-out circuit on the mechanical  system. The current flowing in the
$LC$  loop of  the  circuit in  \fref{sphere_twomode_cap} generates  a
force on  the mechanical resonator proportional to  the current itself
and  the   applied  electric  field.  The   back-action  force  vector
$\mathbf{F_{BA}^R}(\omega)$  adds  to  the Langevin  force  generators
$\mathbf{F_N^R}(\omega)$ introduced in  \eref{Mechmatrix} and is equal
to                                                            
\beq{FBA}
\mathbf{F_{BA}^R}(\omega)=(\frac{E_1I_{p,1}}{j\omega}.. 
\frac{E_NI_{p,N}}{j\omega})=\frac{\mathbf{E}\mathbf{I_p(\omega)}}{j\omega}.
\eeq

The complete set of coupled equations of motion becomes finally:
\beq{full_eq_of_motion}
\left[\begin{array}{c c }
\mathbf{\cal M} & \begin{array}{c c} \mathbf{Z_{BA}} & \mathbf{0} \end{array}
 \\
\begin{array}{cc} \mathbf{0} & \mathbf{E}\\ 
       \mathbf{0} &\mathbf{0} \end{array} & \mathbf{Z}
\end{array}\right]\left[ \begin{array}{c}
      \mathbf{a}(\omega)\\
      \mathbf{q}(\omega)\\
      \mathbf{I_p}(\omega) \\
      \mathbf{I_i}(\omega)
   \end{array} \right]=\mathbf{A'}\left[ \begin{array}{c}
      \mathbf{F_N^S}(\omega) \\
      \mathbf{F_N^R}(\omega) \\
      \mathbf{V_{r}}(\omega) \\
      \mathbf{V_n}(\omega)
   \end{array} \right],
\eeq

\noindent where ${\cal M}= -\omega^2\mathbf{M}+\mathbf{K}$, $\mathbf{Z_{BA}}$ 
is the $(5+N)\times N$ back-action matrix  given by
\beq{BAmatrix}
\mathbf{Z_{BA}}=\left[\begin{array}{c}
-\alpha \mathbf{B}\\
\mathbf{I}\end{array}\right]
\frac{\mathbf{E}}{j\omega},
\eeq

\noindent and 

\beq{Aprime}
\mathbf{A'}=\left[\begin{array}{cc}
\mathbf{A} & \mathbf{0}\\
\mathbf{0} & \mathbf{I}
\end{array}\right].
\eeq

The $(5+3N)$ square matrix on the left side of \eref{full_eq_of_motion} can be
seen as the impedance matrix $\cal Z$ of the electro-mechanical system. 
Defining
$\mathbf{G}=\mathbf{\cal Z}^{-1} \mathbf{A'}$, the SQUID input current for each
transducer is given
by  
\beq{Isqinput}
\mathbf{I_i}=\mathbf{G_I}\left[ \begin{array}{c}
      \mathbf{F_N^S}(\omega) \\
      \mathbf{F_N^R}(\omega) \\
      \mathbf{V_{r}}(\omega) \\
      \mathbf{V_n}(\omega)
   \end{array} \right]=\mathbf{G_IF},
\eeq

\noindent where $\mathbf{G_I}$ is a submatrix of the admittance matrix 
$\mathbf{G}$ with components $G_{s,r}$, where $s=5+2N..5+3N$ and $r=1..5+3N$. 
The vectors ${\mathbf{F^S}}$, ${\mathbf{F^R}}$,${\mathbf{V_r}}$ and 
${\mathbf{V_n}}$ are the forces generated by each noise source.

The noise of the detector, referred to the SQUID amplifier input, in absence of
 signal, is described by the spectral density matrix
$\mathbf{S_I}$ \cite{Helstrom68}. Each component of the matrix, 
\beq{Sinm}
S^{mn}_I=\int^\infty_{-\infty} e^{-j\omega\tau}R_{nm}(\tau) d\tau =\langle
I^m_I(\omega)I^{n*}_I(\omega)\rangle,
\eeq
is the Fourier transform of the correlation function for the $m$th and $n$th
outputs defined as
\beq{corrImIn}
R_{nm}(\tau)=\langle I^m_I(t)I^n_I(t-\tau) \rangle=\int^\infty_{- \infty}
I^m_I(t)I^n_I(t-\tau) dt.
\eeq
From \eref{corrImIn},  each component of the spectral density matrix becomes 
the product of the Fourier transforms of the transducer outputs, as shown in 
the second equality of \eref{Sinm}. The white current  noise $I^n_0$ of the 
SQUID amplifier, which will be better defined below, needs to be added to the 
spectral matrix $S_I$. In matrix notation we can easily write the total SQUID 
current spectral density matrix as 
\beq{Simatrix}
\mathbf{S_I}=\mathbf{G_IFF^*G_I^*}+\mathbf{S_{I,0}}, 
\eeq
where the N square matrix $\mathbf{S_{I,0}}$ has components 
$S^{m,n}_{I,0}=I^m_0I^{n*}_0$. The diagonal elements  of $S_{I,0}$ are equal to 
the current spectral density given in \eref{CGT}. The correlation between the 
$N$ SQUIDs additive  current noise is expected to be negligible so, in the 
following, we will consider diagonal the matrix $\mathbf{S_{I,0}}$.    
The spectral density defined in \eref{Simatrix} can be numerically calculated 
and experimentally measured by means of techniques where the phase information 
is preserved. 

The optimal signal to noise ratio $\rho_0$ , for a gravitational wave signal of
 amplitude $\tilde{h}(\omega)$, is given by
\beq{SNR}
\rho_0^2=4\int^\infty_0 \frac{\tilde{h}^2(\omega)}{S_{hh}(\omega)}
\frac{d\omega}{2\pi},
\eeq
where
\begin{multline}\label{strainnoise}
S_{hh}(\omega)=\mathbf{h}^*(\omega)\left[\mathbf{{(F_m^S)^\ast}}
\mathbf{G_{sig,I}^\ast}\mathbf{S_I^{-1}(\omega)}\mathbf{G_{sig,I}}
\mathbf{F_m^S}\right]^{-1}\mathbf{h}(\omega)\\
=\frac{4}{(m_S \chi R \omega^2)^2}\left[\mathbf{{T_V^\ast}}
\mathbf{G_{sig,I}^\ast}\mathbf{S_I^{-1}(\omega)}\mathbf{G_{sig,I}}
\mathbf{T_V}\right]^{-1}
\end{multline}
is the one-sided total {\it strain} noise power spectrum. 
 
In the equation above,  we called $\mathbf{h}$ the vector $(h_+ h_\times)$ and
the $N\times 5$ matrix $\mathbf{G_{sig,I}}$ is a sub-matrix of the admittance 
matrix $\mathbf{G(\omega)}$ with components $G_{s,r}$, where $s=5+2N+1..5+3N$ 
and $r=1..5$. 
 
Each  transducer line  should be  considered as  a linear  system with
the ($5+2N+2$)  uncorrelated noise  sources described above,
if  we consider  as
negligible the  correlation between the  voltage and current  noise in
the SQUID amplifier.   However, the outputs of the  N transducers do have
correlated  noise  and the  off-diagonal  components  of the  spectral
density matrix are  non zero. One can always  find linear combinations
of   transducers  outputs,  which   produce  N   uncorrelated  signals
\cite{Stevenson98}. Since $S_I$ is  an $N\times N$ Hermitian matrix it
can be  diagonalized by  an unitary matrix  $U(\omega)$. A  new output
channel vector is then obtained  and is related to the original vector
$\mathbf{I_i}$ by \beq{IU} \mathbf{I_i^u}=\mathbf{U^\ast I_i}.  \eeq

The  channels $\mathbf{I_i^u}$ are  statistically independent  and the
spectral  density matrix  \beq{SUmatrix} \mathbf{S_i^u}=\mathbf{U^*S_i
U}=diag(\psi_1(\omega),...,\psi_N(\omega))   \eeq  is   diagonal,  the
eigenvalues  $\psi_i$  of $\mathbf{S_i^u}$  being  the noise  spectral
density of each independent channel.

We   notice   that,    after   performing   the   diagonalization   in
\eref{SUmatrix}, the  total optimal SNR can  be written as  the sum of
the   SNR   in   each   statistically  independent   output   channel.
\beq{SNRsum}
\rho_0^2=\sum^N_{i=1}4\int_{0}^{+\infty}\frac{|I_i^{u,sig}|^2}{\psi_i}
\frac{d\omega}{2\pi}.  \eeq  
From \eref{SNR} and \eref{strainnoise} the SNR becomes 
\begin{multline}\label{SNRburst}           
\rho_{0}^2=\left(\frac{m_{s}\chi R
    \omega_0^2}{2}\right)^2\frac{|\tilde{h} (\omega)|^2}{2\pi}\times    \\
          \quad         \quad          \times
\left[4\int_{0}^{+\infty}\mathbf{T_V^*G_{sig,I}^*(\omega)}
\mathbf{S_I^{-1}(\omega)}\mathbf{G_{sig,I}(\omega)T_V}d\omega\right].
\end{multline}

The performance of resonant detectors is often characterized by
their sensitivity to impulsive burst signals which vary little over
the  detection bandwidth. For impulsive signals, the SNR is
proportional to the deposited energy $E$ in the antenna initially at
rest given by, \cite{MISTHORWHE73}, 
\beq{Edep}
E=\frac{c^3}{G}\frac{1}{16\pi}\omega_0^2|\tilde{h}(\omega_0)|^2\Sigma=\frac{\pi}{4}
\frac{\rho_Sv_S^5}{f_0}\Pi|\tilde{h}(\omega)|^2,
\eeq 
where $\Sigma=\frac{G}{c^3}\frac{\rho v_S^5}{f_0^3}\Pi$ is the integrated cross section of a spherical detector
\cite{WagPaik77} and $\Pi$ is the reduced energy cross section equal
to $0.215$ for a CuAl sphere \cite{ZhouMich95}. $G$ and $c$
are Newton's gravitational constant and the speed of light
respectively, while $\rho_S$ and $v_S$ are the
antenna material density and sound velocity. 

 The pulse detection
noise temperature $T_N$ is then defined as
\beq{Tpulse}
T_N=\frac{1}{k_B}\frac{E}{{\rho_0}^2}.
\eeq

This is a convenient quantity to compare spherical detectors
with bar detectors: while $E$ for a bar-antenna depends on source direction and
polarization, $T_{N}$ does not. 
Using \eref{SNRburst} and \eref{Edep} we can write $T_N$ as follows
\begin{multline}\label{TNeq}  T_N =  \frac{4
 \pi}{k_B}\left(\frac{\pi}{m_S \chi R}\right)^2\frac{\rho_S v_s^5}{\omega_0^5}  \times \\
\quad\times \left[4\int_{0}^{+\infty}\mathbf{T_V^*G_{sig,I}^*(\omega)}
\mathbf{S_I^{-1}(\omega)}\mathbf{G_{sig,I}(\omega)T_V}d\omega\right]^{-1}.
\end{multline}

We remark here that, due to the dependency of the matrix
$\mathbf{T_V}$ from 
the wave direction and polarization, the sensitivity of a spherical detector 
will be isotropic over the sky only if a sufficient number of transducer 
is used ($N>5$). For $N<5$, one can define the detector 
sensitivity by averaging over the direction and polarization as described 
in \cite{Dubath06}.  

\section{Noise contributions}
\label{noisesources}
\subsection{Mechanical   resonators}  Thermal   noise   is  the   main
contribution  of  the  mechanical  resonators to  the  total  detector
noise. The  spectral density of the thermally  activated forces acting
on    the   mechanical    modes    can   be    estimated   from    the
fluctuation-dissipation  theorem  \cite{Saulson90},  and  described  as
follows
 
\beq{mechthermnoise}            \mathbf{S_{F,th}}(\omega)=4k_BTRe\left(
\frac{\cal M}{j\omega},  \right), \eeq where  $\frac{\cal M}{j\omega}$
is    the   system    mechanical   impedance    matrix    derived   in
\eref{full_eq_of_motion}, $k_B$ is the Boltzmann constant and T is the
thermodynamic temperature.   With this formalism we  take into account
the stochastic crosscoupling among the various degrees of freedom of a
macroscopic mechanical body \cite{Majorana97}.

The  thermal  noise  contribution  of  other modes  of  the  spherical
detector  has never  been considered  as  a possible  source of  noise
because they lay generally far a way from the detector bandwidth. In a
real detector,  the first  toroidal modes  are only a  few tens  of Hz
lower than  the spheroidal modes due  to the spread  of the resonances
caused by  the detector  asymmetry \cite{MINIGRAIL04}. Moreover,  in a
spherical  detector  with large  bandwidth,  the  toroidal modes  will
unrecoverably fall into the  sensitive bandwidth. However they are not
sensitive to GW and they couple very weakly to a resonator with radial
sensitivity.  The  toroidal modes  usually have large  quality factors
than  the spheroidal  modes and  their thermal  noise  contribution is
generally negligible.

Further-more,  higher frequency  modes  could also  contribute to  the
total  noise  in  the   detector  bandwidth  due  to  down-conversions
phenomena  related  to the  physical  geometry  and  dimension of  the
read-out transducer  \cite{Levin98}.  To evaluate  the contribution of
higher  frequency  mode  one  can  proceed as  in  \cite{Brian03},  and
calculate  the  total mechanical  impedance  using  Finite Element  or
numerical techniques.  Here we consider  only the thermal noise of the
spheroidal modes and the main radial resonance of the transducers, the
latter being  usually the dominant source of  noise.  The contribution
from other modes will be studied in a following paper.

\subsection{Electrical resonators} The LC resonator which derives from
transducer  capacitance   and  the   primary  coil  of   the  matching
transformer  contributes to  the total  noise with  a  thermal voltage
noise source  associated with the resonator  losses, with single-sided
spectral density  $S_V=4 k_B T Re  (Z_{LC}(\omega))$.  The dissipating
term  $r=Re(Z_{LC}(\omega))$  is linked  to the  intrinsic electrical
quality  factor  of  the  LC  resonator  by  the  well-known  relation
$Q_{el}=\omega_{res}L_p/Re  (Z_{LC})$. The dissipation  resistance $r$
includes the  contributions from  dielectric losses in  the transducer
and decoupling capacitor, in the coil parasitic capacitance and in the
coil  insulating layers and  holder \cite{Falferi94},  magnetic losses
due to flux motion in the superconducting shields, "magneto-resistive"
losses  due   to  dissipative   components  in  the   SQUID  amplifier
\cite{Falferi03}.

The thermal noise contribution from the LC resonator adds to the SQUID
back-action  noise and  may become  significant when  SQUID amplifiers
with $\epsilon < 200 \mathrm{\hbar}$ are employed.

\subsection{SQUID amplifier noise theory}

A  coupled dc  SQUID amplifier  can be  modelled as  an  ideal current
amplifier with a current noise  source $I_n$ in parallel and a voltage
noise source $V_n$ in series with  the input coil. The two of them are
responsible respectively for additive  and back action noise. To fully
characterise  the SQUID  it is  necessary to  estimate both  the noise
contributions.  An  useful parameter  to characterize an  amplifier is
its noise temperature $T_n$, defined by
\beq{SQUIDTn}    
T_n=\frac{\sqrt{S_{vv}S{ii}-Im    (S_{iv})^2}}{4k_B}=
\frac{\omega}{k_B}\sqrt{\epsilon_{ii}\epsilon_{vv}-\epsilon_{iv}^2},
\eeq 
in the classical limit when $k_B T_N >> \hbar \omega$.
$T_n$  is the  temperature at which  the optimal  input impedance
gives  the thermal  noise power  equal  to the  amplifier noise.   Its
minimum value  for a  linear amplifier is  imposed by  the uncertainty
principle and is given by $T_{N}sim \hbar\omega/k_B$ \cite{Caves82}.
$S_{vv}$  and $S_{ii}$  are the  spectral densities,  referred  to the
SQUID input coil,  of the two noise generators,  while $S_{vi}$ is the
cross correlation between the two.  According to Clarke-Tesche-Giffard
(CTG) theory the single-sided spectra are equal to
\beq{CGT}
\begin{array}{l}      S_{ii}\simeq      16      \frac{k_B      T}{R_s}
\left(\frac{L_{SQ}}{M_{i,SQ}}\right)^2 \\ \\ S_{vv}\simeq 11 \frac{k_B
T}{R_s} \omega  (M_{i,SQ})^2 \\  \\ S_{jv}\simeq 12  \frac{k_B T}{R_s}
j\omega L_{SQ},
\end{array}  \eeq where $R_s$,  $L_{SQ}$, $M_{i,SQ}$  are respectively
shunt resistors,  self inductance and input coil  mutual inductance of
the  SQUID.  The  cross correlation  power spectrum  is  always purely
imaginary  due  to the  time-reversal  simmetry  of  the SQUID  motion
equation. Its absolute  value is usually small and  we will neglect it
in   the  following   calculations.   In   \eref{SQUIDTn}   the  noise
temperature  is   also  defined   in  terms  the   energy  resolutions
$\epsilon_{ii}$,  $\epsilon_{vv}$  and  $\epsilon_{iv}$  expressed  in
units  of  $\hbar$.   In  the  following sections  we  will  not  make
distinctions  between  the  voltage  and  current  energy  resolution,
$\epsilon_{ii}$  and  $\epsilon_{vv}$, and  we  will generically  talk
about SQUID  energy resolution assuming  that they both have  the same
value.  The  cross correlation energy  resolution $\epsilon_{iv}$ will
be neglected.

\section{Numerical analysis}
\label{analysis} Here we  numerically calculate the noise temperature,
the bandwidth  and the strain sensitivity of  an ultracryogenic, large
$2  m$  in   diameter,  spherical  detector  as  a   function  of  the
transduction  chain  parameters.   We  consider a  detector  operating
according to the state-of-art resonant antenna technology and when all
the  parameters are  improved to  operate the  detector nearly  at the
quantum  limit.   Further we calculate the antenna patterns and the 
sky coverage
of  two  identical  detectors  operated  in  coincidence  and  located
respectively  in  Leiden  (The  Netherlands) and  Sa$\tilde{o}$  Paolo
(Brasil), the  location of the  two small spherical  antenna Minigrail
\cite{MINIGRAIL}, and Mario  Schenberg \cite{SCHENBERG}, currently under
development.  Finally the anisotropy  in the sensitivity and bandwidth
is studied for a not ideal resonator and a not optimally tuned 2-mode
capacitive transducer.

\subsection{Parameters optimization}  To optimize a  resonant detector
one must  know the  voltage and current  noise of the  available SQUID
amplifier. The  other parameters can  then be adjusted to  achieve the
best  SNR.   The optimal  impedance  matching  between the  mechanical
resonators  and the SQUID  amplifier is  achieved when  the transducer
electrical mode is tuned to  the mechanical resonances.  In the scheme
described   in    \fref{sphere_twomode_cap},   the   electrical   mode
contributes  to   the  transducer   chain  with  an   equivalent  mass
$m_{eq}=C_TE^2/\omega_{el}^2$.   In a  multimode  detector the  energy
transfer     between    each     resonator    is     optimized    when
$\mu=m_R/m_{eff}=m_{el}/m_R$, where $\mu$ is defined as the mass ratio
and  $m_{eff}$ is  the effective  mass  ratio of  the five  spheroidal
modes, which is a fraction of the sphere total mass $m_s$ and is equal
to  $m_{eff}=5/6\chi m_s$  \cite{Merkowitz95}. For  optimal mass-ratio
one gets a shorter energy transfer  time between the modes and a wider
system  bandwidth.  To fully  describe the  sensitivity of  a resonant
detector  we  consider  the   signal-to-noise  ratio  SNR,  the  pulse
detection  noise  temperature $T_N$  and  the  signal bandwidth,  here
described       by        \cite{vantrees}:       \beq{BW}       \delta
f=\frac{1}{2\pi}\frac{\left(\int_0^\infty         |S_{hh}^{-2}(\omega)|
d\omega\right)^2}{\int_0^\infty    |S_{hh}^{-2}(\omega)|^2   d\omega}.
\eeq

In   the  analysis   below  we   imposed  the   following  conditions:
\beq{conditions}
\begin{array}{l} E=\mu\omega_R \left(\frac{m_{eff}}{C_T}\right)^{1/2},
\\ \\ \omega_{el}^2=\frac{1}{C_TL_r}=\omega_{tr}^2,
\end{array}  \eeq where  $L_r=L_p(1-k^2L_s/(L_s+L_i))$ is  the reduced
inductance  of  the primary  coil  of  the  matching transformer  with
coupling  $k^2=M^2/(L_sL_p)$.   The  first  condition is  obtained  by
assuming  a constant  mass  ratio $\mu=m_{el}/m_R$.   With the  second
condition we  impose that the  electrical resonances are tuned  to the
transducers  mechanical resonances. This  condition is  fundamental to
get optimal matching in a capacitive transducer. The parameters of the
electrical matching  network depend  on the value  of the  SQUID input
inductance and noise, and on the  detector mass ratio. It can be shown
that for large mass ratio, a coupling $k\sim 1$, and optimal matching,
the inductance $L_s$ of the superconducting transformer secondary coil
should be  as large as the  SQUID input inductance  $L_i$.  However in
practice  large $L_s$  will  guarantee large  coupling  and finally  a
better  sensitivity. For  low mass  ratio, the  impedance seen  by the
transformer primary  coil is lower than  for high mass  ratio when the
same capacitance  is considered. One  can obtain a better  matching by
decreasing both  the inductances  $L_p$ and $L_s$  of the  primary and
secondary    coil    while    maintaining   valid    the    conditions
\eref{conditions}.  The  improvement is however  rather limited giving
for example about $20\%$  better noise temperature using an inductance
$L_s\sim 0.3L_i$ for a mass ratio $\mu=0.001$ and a $k\sim 1$.
 
We anticipate that for a detector with significant thermal noise (high
$T/Q$ ratio) operating  with a low noise SQUID  amplifier, the optimal
sensitivity is obtained for  large mass ratio $\mu$ and, consequently,
high  transducer capacitance (see  \eref{conditions}).  It  is however
physically  impossible  to reach  arbitrary  resonator  high mass  and
capacitance  using  the detector  configuration  discussed here.   The
maximum transducer capacitance $C_T$ and mass $m_R$ one can reasonably
obtain, which however have never  been developed so far, are estimated
to  be $C_{T,max}\simeq  30  nF$ and  $m_{R,max}\simeq  90 Kg$.   They
correspond  to a  transducer with  an area  $A\simeq 0.1  m^2$,  a gap
$d\simeq 20  \mu m$  and a mass  ratio $\mu \simeq  0.01 $\footnote{To
obtain large  capacitance one could for  example make use  of a double
electrode scheme as  proposed in \cite{Bassan95}. However, transducers
with such  a large  mass and  capacitance and high  Q have  never been
developed so far  and the feasibility needs to  be demonstrated.}.  We
notice that  strong physical constrains  apply in general also  to the
mechanical  resonant frequency  due  to the  limits  on the  resonator
membrane thickness \cite{Solomonson92,Bassan95}.  However, a resonator
"rosette" design as developed  for the NAUTILUS and EXPLORER detectors
\cite{Bassan95},  allows a  rather wide  freedom in  the choice  of the
resonator mass and sensitive area for a give resonant frequency.

We report  below the  calculated detector  sensitivity for a  $ 2  m $
large in  diameter, 30 ton, CuAl  sphere.  The main  parameters of the
detector are  summarized in  \tref{spheretable}.  Such an antenna has
a cross section $\Sigma=9.76\cdot 10^{-24}m^2Hz$.
 The total energy deposited by a GW of
amplitude $\tilde{h}(\omega)$ is $E=1.0\cdot
10^{35}\omega^2|\tilde{h}(\omega)|^2\, K$. 
A GW burst signal, lasting for a time $\tau_G~1ms$  shorter than the 
detector integration time  and rising
quickly to an amplitude $h_0=\tilde{h}(\omega_0)/\tau_G\sim 2\cdot
10^{-21}$, deposits an energy $T_{GB}=1\mu K$.

We consider  a sphere
equipped   with   six  radial   capacitive   resonators   in  the   TI
configuration.   The electrical mode  of the  superconducting matching
network is tuned to the mechanical ones. The signal of each transducer
is amplified by a two-stage SQUID amplifier.
\begin{table}[htbp]
  \begin{center} \small
    \begin{tabular}{c c c}  
\hline  {\it } &\multicolumn{1}{c}  {\it current}
&\multicolumn{1}{c} {\it quantum}  \vspace{0.05 cm}\\ {\it parameters}
&\multicolumn{1}{c}  {\it technology} &\multicolumn{1}{c}  {\it limit}
\vspace{0.05 cm}\\ \hline  sphere mass , $M_s \,[tons]$&  $ 30$ & $30$
\\  sphere diameter  $\; [m]$&  $  2$ &  $2$ \\  resonator mass,  $M_r
\,[Kg]$& $  10 $ &  $90 $ \\  spheroidal modes, $f_l \,  [Hz]$& $
987, 1001,1008,$ & $987, 1001, 1008,$ \\ & $1012, 1017$ & $1012, 1017$
\\  $Q_s$,$Q_r$,$Q_{el}$ & $2\cdot  10^6 $  & $5\cdot  10^7 $\\  $T \;
[mK]$  &$50$ &$20$ \\  $C_t \;  [nF]$& $10$&  $30$ \\  $L_p \;  [H]$ &
$3.6$& $1.2$  \\ bias  field, $E \,  [Volt/m]$ &  $5 \cdot 10^6$  & $4
\cdot 10^7$\\  SQUID sensitivity, $E_{res}\,  [\hbar]$ & $50$  & $1$\\
\hline
     \end{tabular}
     \caption{ Main  parameters used in the numerical  analysis of the
sensitivity  of  a spherical  detector  equipped  with six  capacitive
transducers. The first column shows the  parameters of a 2 m sphere in
CuAl already achieved in separate experiments. In the second column we
give the parameters  necessary to operate the detector  at the quantum
limit.  The  CuAl alloys  have a  sound velocity of  $4700 m/s$  and a
Poisson  ration of about  0.3.  The  material dependent  factor $\chi$
introduced in  \eref{effforce}is $\chi=0.327$ for  CuAl. The effective
mass  of the  spheroidal  modes is  then $m_{eff}=5/6\chi*m_s\simeq  8\, 
tons$.}
     \label{spheretable}
  \end{center}
\end{table}
 Two  situations  have been  considered.   In  the  first  one  the  detector
parameters   have   values   according   to  the   current   available
technology. In the second we  estimate the ultimate sensitivity of the
detector obtainable  with quantum limited amplifiers and  ultra high Q
mechanical  and electrical resonators.   For a  given $T/Q$  ratio and
SQUID amplifier energy resolution, the detector sensitivity depends on
the  resonators   mass  ratio,  the  electric  field   bias  $E$,  the
transducers   capacitance   and  the   parameters   of  the   matching
transformers.

\begin{figure}
 \includegraphics[scale=1.1]{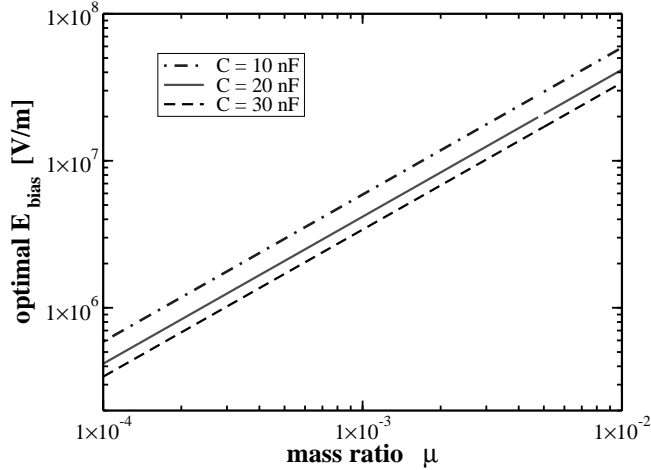}
 \caption{\label{ECmu} Optimal bias electric field for different value
of the  transducer capacitance  as a function  of the  resonators mass
ratio $\mu$.}
\end{figure}

 The detector  effective noise temperature, the bandwidth  and the SNR
for a  GW burst of $T_{GB}=10\mu  K$ are calculated as  a function of
the mass ratio  for different values of the  $T/Q$ ratio. $T/Q$ refers
both to the mechanical  and electrical resonances. The equivalent mass
of the electrical resonator is adjusted according to the mass ratio by
changing the  electric field bias  E. The optimal electrical  field is
calculated  as  a  function  of  the mass  ratio  and  the  transducer
capacitance. The results are shown in \fref{ECmu}.
We considered mechanical  and electrical $T/Q$ of $1\cdot
10^{-7}$, $2.5\cdot 10^{-8}$,  $4\cdot 10^{-10}$.  Mechanical $T/Q$ of
the order  of $10^{-8}$ have been  reached in a CuAl  sphere cooled at
$50  \mathrm{mK}$ \cite{deWaard04}, and  in Al5056  bars cooled  at 100
mK. The lowest $T/Q=3\cdot 10^{-8}$ for electrical resonators has been
achieved by  the AURIGA  group with a  large Nb coil  resonator cooled
down  to  $50   \mathrm{mK}$  \cite{Falferi06}.   The  fabrication  of
electrical   resonators   with   $T/Q=1\cdot  10^{-7}$   at   acoustic
frequencies  is  nowadays  a  well established  technology.   A  $T/Q<
10^{-8}$ has not yet been achieved experimentally either in mechanical
transducers or in electrical resonators.
\begin{figure*}
 \includegraphics[scale=1]{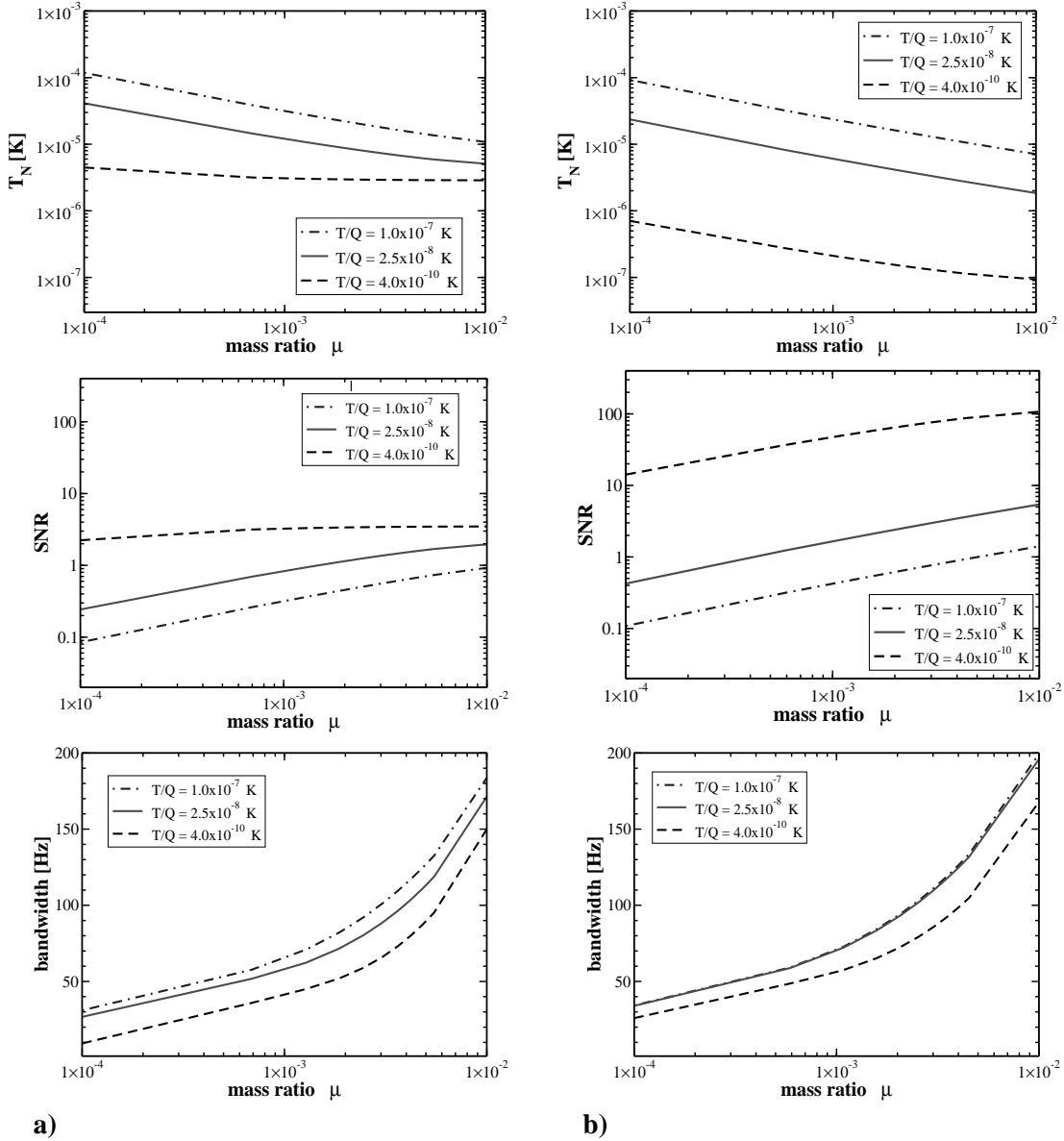}
 \caption{\label{optimum} Noise temperature,  SNR and bandwidth of the
spherical detector  as a function  of the resonator mass  ratio $\mu$.
The  detector   is  read-out  by   a  $50  \hbar$   energy  resolution
$\mathbf(a)$ and quantum limited  $\mathbf (b)$ SQUID amplifier. Three
different   configurations  with   $T/Q=1\cdot   10^{-7}K$,  $2.5\cdot
10^{-8}K$  and  $4\cdot 10^{-10}K$  are  studied. The SNR is calculated for a
$T_{GB}=10\mu K$ gravitational wave burst.}
\end{figure*}  
The results  of  the calculations  are  shown in  plots
\fref{optimum}$\mathbf{a)}$, where the  noise temperature, the SNR and
the bandwidth are given as functions  of the mass and the $T/Q$ ratio.
The  SQUID  amplifier  coupled  energy  resolution was  chosen  to  be
$\epsilon_{coupled}=50 \hbar$.  A coupled  energy resolution as low as
$27 \hbar$ has been recently obtained with a two-stage SQUID amplifier
coupled  with  an  electrical  resonator  and  cooled  down  to  $T=50
\mathrm{mK}$ \cite{Falferi06}. The  spherical GW antenna Minigrail and
the  AURIGA  detector  are  currently  operating at  5  K  with  SQUID
amplifier   energy   resolution   of   the  order   of   $600   \hbar$
\cite{GottardiPhD,Baggio05}. More  than a factor of  10 improvement is
expected when this detector will operate at $T<100 mK$.

From  the plots  in  \fref{optimum} we  see  that, in  terms of  noise
temperature and  bandwidth, a large mass  spherical detector developed
using  the  available  technology  will  not perform  better  than  an
ultra-cryogenic bar  detector operating at  the same frequency.   As a
matter of fact, the noise  temperature $T_N$ and the bandwidths depend
only on  the frequency, the electro-mechanical  impedance matching and
the SQUID amplifier noise. However, due to the larger cross section of
a  spherical detector,  with respect  to a  bar detector  at  the same
frequency,  the spherical  detector improves  the SNR  by a  factor of
about  40  \cite{ZhouMich95}.   Moreover, omnidirectionality  will  of
course still be a unique feature of a spherical detector when at least
six resonators are used. With  $T/Q= 2.5\cdot 10^{-8}K$ and mass ratio
$\mu=0.001$, corresponding to a transducer resonating mass of about $9
\,  Kg$, a  capacitance of  $10 nF$  and an  optimal  electrical field
$E=5\cdot 10^6  V/m$, the  detector has a  noise temperature  $T_N$ of
about $10\mu  K$, about a  factor of 30  better than the  present most
sensitive resonant bar detector \cite{Baggio05}.

The SQUID  amplifier is considered strongly coupled  to the electrical
matching  circuit  assuming  $k=  0.8$,  $L_p\sim 1  H$  and  $L_s\sim
L_i=1.7\mu H$.   As in bar  detectors, the sensitivity will  be mainly
limited by thermal noise of the electrical resonator, and, outside the
resonances,  by the  SQUID amplifier  additive noise.   The  curves in
\fref{optimum}.a clearly  show that  large mass ratio  brings benefits
only to the bandwidth and  not to the detector noise temperature. This
is  due to  the fact  that a  large mass  resonator will  not  help to
decrease the  thermal noise  contribution.  As a  matter of  fact, the
contribution of  the mass at  the denominator of the  Langevian forces
derived in  \eref{mechthermnoise} is almost cancelled  by the transfer
function when transforming the  forces into displacement. On the other
end, the bandwidth will increase for  large mass ratio due to a better
impedance   matching.   The  latter   will  unavoidable   enlarge  the
contribution of the back action  noise coming from the SQUID amplifier
producing the observed saturation in the detector sensitivity at large
mass ratio.

\noindent  To fully exploit  the potentiality  of a  massive spherical
detector  one  needs  to  develop  capacitive  transducers  with  high
sensitive  area, massive mechanical  resonators with  Q-factors larger
than $10^7$,  high Q electrical  transformers and large  bias electric
fields. 

In \fref{optimum}.b the detector  sensitivity is given for the antenna
and  transducer chain  parameters  obtainable by  pushing the  current
technology  to  its  limits.   We  consider a  quantum  limited  SQUID
amplifier operating  at temperature $T<100 mK$. Again  we estimate the
noise temperature  $T_N$, the bandwidth and  the SNR for a  $10 \mu K$
burst for $T/Q=1\cdot  10^{-7}K,\, 2.5\cdot 10^{-8}K,\, 4\cdot 10^{-10}K$.
By  comparing  those plots  with  the  ones  previously discussed,  it
becomes  clear  that without  an  improvement  of  the mechanical  and
electrical resonators with respect to available technology, the use of
a  quantum  limited SQUID  amplifier  will  not  benefit the  detector
sensitivity. By  looking at  the dot-dashed curves  of \fref{optimum},
for  examples, corresponding  to  a $T/Q=1\cdot  10^{-7}K$, one  shall
expect no improvement in the strain sensitivity, but only a small increase
of the  bandwidth due  to the  lower additive noise  of the  SQUID.  A
spherical  detector  operating  at  $T=20\, mK$  with  mechanical  and
electrical  quality factor  $Q\sim 5\cdot  10^{7}$, a  quantum limited
SQUID amplifier and large  mass mechanical transducers, $\mu=0.01$ and
$m_R=90\,Kg$,  can have  a noise  temperature of  $1.3\cdot 10^{-7}K$,
corresponding to a peak  strain sensitivity of $10^{-23} Hz^{-1/2}$ at
$1kHz$, and  a bandwidth of about  $200 Hz$.  This would  improve of a
factor 50 the  sensitivity of an existing bar  or small sphere antenna
working at the quantum limit.  The bandwidth will have only a moderate
increase   with    respect   to   present    resonant   bar   antennae
\cite{Baggio05}, merely due to  the spreading of the spheroidal modes.
The  minimum achievable  antenna  noise temperature  is  given by  the
quantum-mechanical  limit  of a  linear  motion  detection derived  by
Giffard         \cite{Giffard76}          and         equal         to
$T_{N,min}=2\hbar\omega/k_B[exp(\hbar   \omega/k_BT_{N,a})-1]$,  where
$T_{N,a}$ is  the noise temperature  of the linear amplifier.   When a
quantum limited amplifier  is used, $T_{N,min}=9.5\cdot10^{-8}K$ for a
kHz resonant sphere considered here. 

\subsection{Strain sensitivity}

We    show   here    the   strain    sensitivity    calculated   using
\eref{strainnoise}.   The strain  curves  are derived  for each  noise
contribution  described in  \sref{noisesources}. The  read-out circuit
was optimized as discussed above.

\begin{figure}
 \includegraphics[scale =1.1]{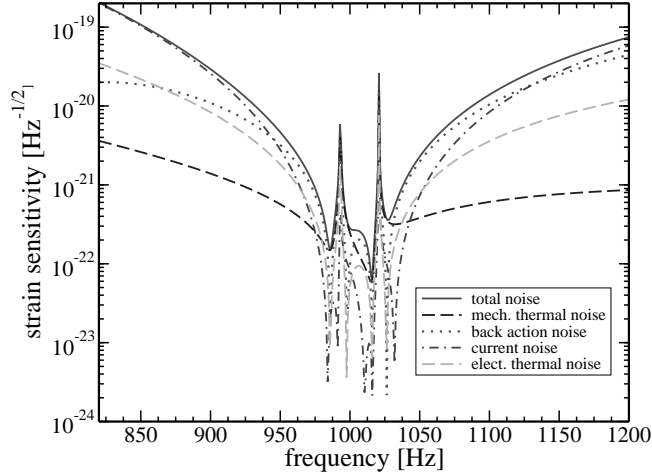}
 \caption{\label{SSh50h1tr}   Strain  sensitivity   for   a  spherical
detector  with  one single  transducer,  $T/Q\sim 2.5\times  10^{-8}$,
resonators  mass  ratio $\mu=0.001$  and  $50\hbar$ energy  resolution
SQUID  amplifiers. The  transducer  electrical mode  is  tuned to  the
mechanical ones. The parameters  used for this simulation have already
been achieved  in separate experiments  with bar detectors or  a lower
mass spherical antenna.}
\end{figure}

\noindent In \fref{SSh50h1tr} the strain sensitivity is calculated for
a spherical detector  equipped with a single transducer  placed in the
position 1.   It was calculated  for an optimally oriented  source, as
will be  described in detail  below. The detector parameters  used for
this  simulation have  already been  achieved in  separate experiments
\cite{Baggio05,EXPNAU,MINIGRAIL}. The sensitivity curves
are obtained considering the detector operating at $T=50mK$ with a $50
\hbar$ SQUID  amplifier and  with the electrical  mode coupled  to the
mechanical ones. The quality factors are chosen to be $Q=3\times 10^6$
and  $Q=2\times   10^6$  for  the  mechanical   and  electrical  modes
respectively.   The  mass ratio  is  $\mu=0.001$,  corresponding to  a
mechanical  resonator mass  $m_R\simeq 9  Kg$. The  optimal electrical
bias  field  is  about   $E=5\cdot  10^{6}  V/m$  and  the  transducer
capacitance is $10 nF$.  

\begin{figure}
 \includegraphics[scale =1.1]{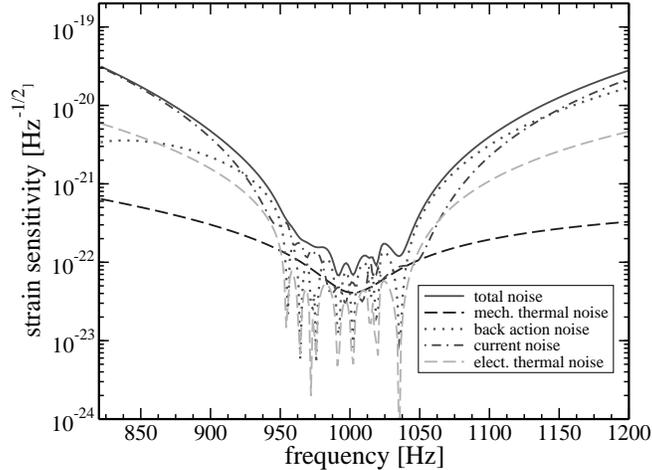}
 \caption{\label{SSh50h} Strain  sensitivity for a  spherical detector
with  six transducers,  $T/Q\sim 2.5\times  10^{-8}$,  resonators mass
ratio    $\mu=0.001$   and    $50\hbar$   energy    resolution   SQUID
amplifiers. The electrical modes are tuned to the mechanical ones. The
parameters  used for  this simulation  have already  been  achieved in
separate  experiments  with  bar  detectors or  lower  mass  spherical
antenna.}
\end{figure}

The strain sensitivity shown in \fref{SSh50h}
was calculated for a spherical detector with six transducers in the TI
arrangement. The detector and  transducers parameter are the same used
in the  single transducer configuration  described above.  Differently
from   the  single  transducer   configuration,  the   sensitivity  is
independent  of  the  wave  direction and  polarization.   In such a 
configuration the detector sensitivity will
be limited by  the thermal noise contribution from  the mechanical and
electrical  resonators and  by  the  back action  noise  of the  SQUID
amplifier due  to the the optimal  matching coming from  the tuning of
the  electrical  resonances with  the  mechanical  ones.  Outside  the
resonances, the  sensitivity is limited by the  additive current noise
of the SQUID amplifier.
\begin{figure}
 \includegraphics[scale =1.1]{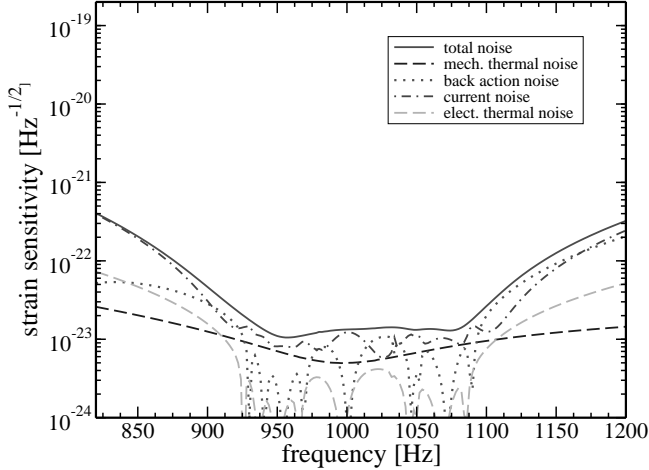}
 \caption{\label{SSh1h}  Strain sensitivity  for a  spherical detector
with  $T/Q=4\times  10^{-10}$, resonators  mass  ratio $\mu=0.01$  and
$1\hbar$ energy resolution SQUID  amplifiers. The electrical modes are
tuned to the mechanical ones.}
\end{figure}

 In  \fref{SSh1h} the  strain sensitivity  is shown  for  an optimized
detector  equipped  with SQUID  amplifiers  operating  at the  quantum
limit. The simulation considers  a detector operating at $T=20mK$ with
all resonators  having quality factors $Q=5\times  10^7$.  We consider
six transducers  in the TI  arrangement. The mass ratio  is $\mu=0.01$
and the optimal electrical bias field is $E\sim 3\cdot 10^{7} V/m$. We
consider a transducer with a large capacitance of about $C=30nF$.  The
sensitivity is mainly limited by additive and back action noise of the
SQUID linear amplifiers, whose minimum energy resolution is imposed by
quantum mechanics. Despite the low  working temperature and the high Q
considered,  a small  contribution  from the  thermal  noise is  still
present.

\subsection{Antenna pattern}

 We calculate here
the  SNR of  a spherical  detector equipped  with one,  three  and six
resonators  as a  function of  the direction  and polarization  of the
incident wave.   We consider  a detector operating at nearly the quantum limit.
 \begin{figure*}
 \includegraphics{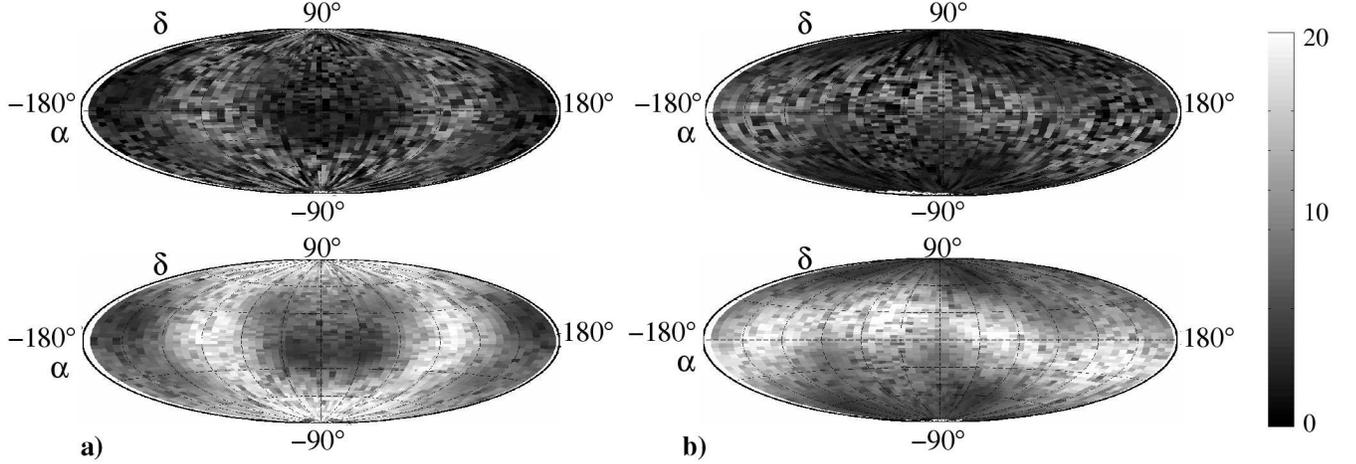}
 \caption{\label{1trCalpharand}   Antenna    pattern   for   spherical
detectors  respectively located  in Leiden  ${\bf a)}$  and  Sa$\tilde{o}$ Paulo
${\bf b)}$ with a single transducer on position 1  of  the six  
positions of the  TI arrangement. 
We  consider a linearly polarized  (top)
and a one-cycle, circularly polarized  (bottom) signal  as explained 
in the text, depositing an energy $T_{GB}=3\mu K$ in the detector.}
\end{figure*}
\begin{figure*}
 \includegraphics{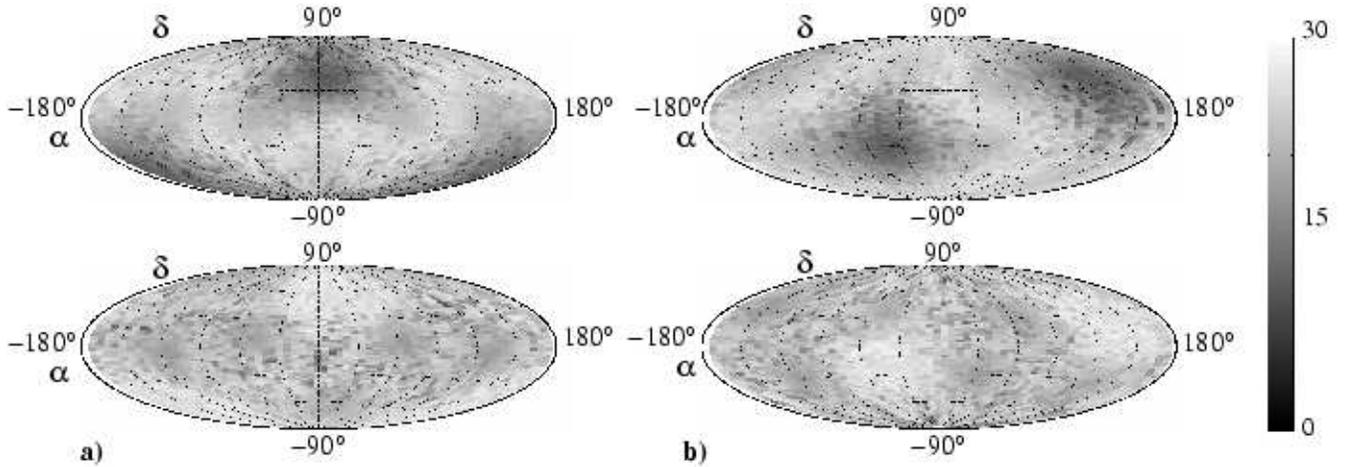}
 \caption{\label{3trCalpharand} Antenna pattern of spherical detectors
respectively  located in  Leiden  ${\bf a)}$  and Sa$\tilde{o}$  Paulo
${\bf b)}$  with a  three transducers set  in the TI  configuration at
$\theta_{TI}=37.3773^{\circ}$  (sky maps above)  and $\theta_{TI}=78.1876^{\circ}$
(sky maps below).  We consider  a one-cycle circularly polarized
sinusoidal signal  depositing an energy  $T_{GB}=3\mu K$ in the detector}
\end{figure*}
We numerically calculate the  SNR as a function of the three
Euler  angles $( \psi,  \theta ,  \phi)$ of  \eref{TV}. All  the three
angles are  necessary to completely  define the GW source.   The first
Euler  angle $\psi$  carries information  about the  wave polarization
\cite{MerkLobo99}.  The  other two, $(\theta ,\phi)$,  give the source
direction. In  the sky maps presented  in this section,  we call those
angles  respectively declination  $\delta=\pi/2-\theta$  and right ascension
$\alpha=\phi$  to  match   the  astronomical  notation.   The  antenna
patterns  depend on  the longitude  and the  latitude of  the detector
location as  well as on  the Universal Time  (UT), $\tau$, due  to the
earth  proper  rotational  motion.   To  simplify  the  discussion  we
consider here detectors at $\tau=0$.
We calculate the  antenna patterns for two spherical  detectors, 2m in
diameter, located  at $Leiden$  (The Netherlands), $lon=4^{\circ}\;  30" ,\;
lat=52^{\circ}\; 7"$, and at Sa$\tilde{o}$ Paulo (Brasil), $lon=-46^{\circ}\; 38", \;
lat=23^{\circ}\;  34"$,  where  respectively  the MiniGRAIL  and  the  Mario
Schemberg,  65 cm  large in  diameter,  spherical  detectors are  being
developed.   We chose  those locations  because the  general discussion
about direction  sensitivity is independent  on the detector  size and
mass a part from the absolute value of SNR and bandwidth. The analysis
presented here is then useful for the  existing small spherical
detectors as well.   

In  \fref{1trCalpharand}  the  sky  maps  are shown  for  a  spherical
detector  with a  single transducer  on  position 1 of the  TI
arrangement.  We consider  a linearly polarized wave with $h_+=h_0$ and
$h_{\times}=0$ and a random polarization angle $\psi$  (skymaps  at
top). The amplitude $h_0$ is related to the
 deposited energy $E=k_B T_{GB}$ according to \eref{Edep}. For each
 simulation we indicate the GW energy in the figure captions.
 The SNR for each wave direction depends on the polarization
angle. The choice of a random polarization for each
direction trial produces the scattered pattern.  When a circularly 
polarized wave is chosen 
 (skymaps at bottom) the maps become smoother and the SNR is only
 dependent on the wave direction. Here and in the following we
 consider a one-cycle,
 circularly polarized  sinusoid: $h_+=h_0\sqrt{2}cos(\omega t)$ and
 $h_\times
=h_0\sqrt{2}sin(\omega t)$, $0 \leq \omega t \leq 2\pi$, with the
frequency $\omega$ laying within the detector bandwidth. 
   Transducers  on  other
positions  will  show  a  similar  pattern rotated  of  proper  angles
accordingly to  the positions. The  sky maps on  the left (\fref{1trCalpharand}.a) refers  to a
detector in Leiden  with the lab-frame oriented so  that the z-axis is
pointing to the local vertical and  the x axis to the local south. The
sky maps on the right (\fref{1trCalpharand}.b) refers to the a detector in Sa$\tilde{o}$ Paulo.
As expected, in  a spherical detector operating with  only one transducer
the sensitivity  is direction and polarization dependent  and changes
 according  to the transducer  location. A  source emitting  a  
linearly or circularly polarized wave is  more likely  to  be detected
when  laying  on the  plane
perpendicular to  the transducer axis  and passing through  the sphere
center.

\begin{figure}
 \includegraphics[scale=0.9]{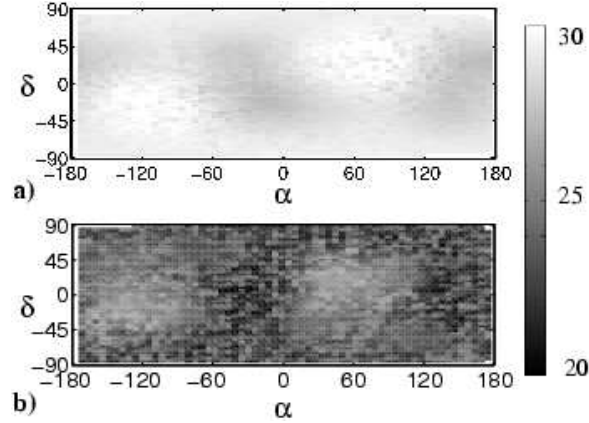}
 \caption{\label{6trSNR} SNR  of a quantum  limited spherical detector
with  six  transducers in  the  TI  configuration.  In  $\mathbf{a)}$,
identical transducers  are considered. In  $\mathbf{b)}$, we arbitrary
modified the parameters of each transducer of a maximum of $15\%$ from
their  optimal value.  We  consider a one-cycle  circularly polarized  wave 
with energy $T_{GB}=3\, \mu K$.}
\end{figure}

\begin{figure}
 \includegraphics[scale=0.9]{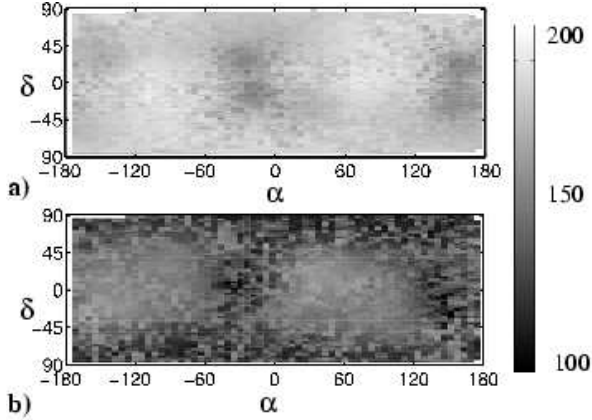}
 \caption{\label{6trBW}   Signal  bandwidth   of  a   quantum  limited
spherical detector  with six transducers  in the TI  configuration. In
$\mathbf{a)}$, identical transducers are considered. In $\mathbf{b)}$,
we arbitrary modified  the parameters of each transducer  of a maximum
of $15\%$ from their optimal value. We consider a one-cycle circularly
polarized sinusoidal signal with energy $T_{GB}=3\, \mu K$.}
\end{figure}

\noindent In \fref{3trCalpharand},  the sky maps
are shown  for a spherical  detector with three transducers  placed at
the   three  positions   of  the   TI  arrangement,   respectively  at
$\theta_{TI}=37.3773^{\circ}$ and  $\theta_{TI}=78.1876^{\circ}$. The
signal used for the simulation is a one-cycle circularly polarized
sinusoid with energy $T_{GB}=3\, \mu K$. The higher maximum SNR with respect
to the single transducer configuration is mainly due to the large
detector bandwidth. 
When  six   identical  resonators  are  used,   the  detector  becomes
independent  form   the  wave  incoming   direction  and  polarization
\cite{MerkLobo99}. This result is also shown in \fref{6trSNR}.  
For  a real detector with  identical transducers,
not optimally tuned to the 5 spheroidal modes, a
small residual anisotropy in the SNR and bandwidth is still present.  
This is however less than $10\%$ over
the  whole   sky.   The  signal  bandwidth  anisotropy   is  shown  in
\fref{6trBW}. Both the figures will be further discussed in the
following section. 

\begin{figure}
 \includegraphics{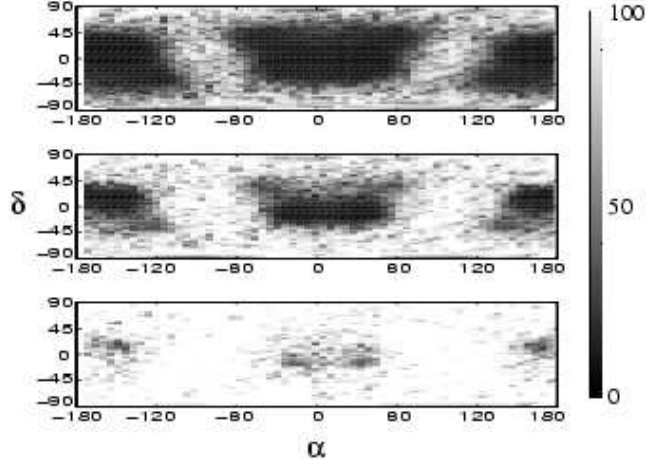}
\caption{\label{skymap_deteff_SNR}  Comparison of detection efficiency sky
  maps for a spherical detector with one transducer and for different deposited energy  $T_{GB}=1,\, 1.5, \, 3 \, \mu  K$ (from top
  to bottom). We consider a circularly polarized wave as described in
  the text.}
\end{figure}

\begin{figure}
 \includegraphics{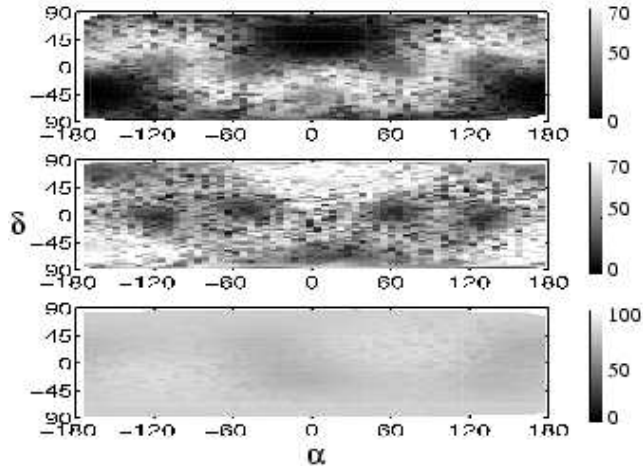}
\caption{\label{skymap_deteff_tr}  Comparison of detection efficiency sky
  maps for a spherical detector with three  transducers in the TI
  positions  respectively at $\theta_{TI}=37.3773^{\circ}$ and  
$\theta_{TI}=73.1876^{\circ}$, and for a complete
detector with 6 transducers (from top to bottom). We consider a GW
burst deposited energy of $0.6\, \mu K$ and a circularly polarized wave. 
Note the difference in the color code
on the graphs and that the signal amplitude is smaller than the ones 
in \fref{skymap_deteff_SNR}.}
\end{figure}

\begin{figure}
 \includegraphics{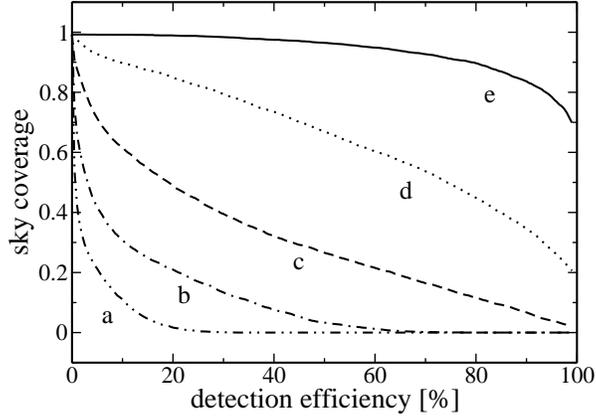}
\caption{\label{skyfrac_deteff}  Fraction of sky as a function of
  detection efficiency for a detector with a single transducer.
 We consider a circularly
  polarized wave with $T_{GB}= 0.6,\,0.75,\, 1,\, 1.5, \, 3 \, \mu  K$, and $\rho_{max}\sim 6, \, 7.5,\, 10,\, 15, \, 30$,  respectively denoted
  by {\bf a}, {\bf b}, {\bf c}, {\bf d}  and {\bf e}.}
\end{figure}

\Fref{skymap_deteff_SNR} presents the detection efficiency of a
spherical detector equipped with only one transducer as a
function of the GW incoming direction for different GW energies,
 $T_{GB}= 1,\, 1.5, \, 3 \, \mu  K$  equal to a maximum SNR 
 respectively  of $\rho_{max}\sim 10,\, 15, \, 30$. In analogy with
 \cite{Arnaud04}, the detection efficiency is defined as
 $\frac{1}{2}{\it erfc}[(\eta-\rho_0)/\sqrt{2}]$, where $\eta$ is a
 threshold chosen equal to 5 and ${\it erfc}$ is the complementary
 error function.
The detection of a one-cycle circularly polarized wave of energy
$T_{GB}=3\, \mu K$ is almost always succesfull independently of the
incoming direction. 
In \fref{skymap_deteff_tr} similar skymaps  are shown for a $T_{GB}= 0.6\, \mu K$ when a spherical detector with 3 and 6 transducers is 
considered.
The results of the skymaps in \fref{skymap_deteff_SNR} are summarized in 
\fref{skyfrac_deteff} where the fraction of sky is plotted as a function of the
detection efficiency for a single transducer configuration and
different signal amplitudes. One notes that the fraction of sky
covered  decreases when the detection probability level increases: the
curves evolution for each SNR can be understood from the patterns of
the skymaps in
\fref{skymap_deteff_SNR}. The detection probability is higher than
$30\%$ in $40\%$ of the sky for a $\rho_{max}=10$, corresponding to a
$1\, ms$ GW burst of amplitude $h_0=2\cdot 10^{-21}$; it is $50\%$ for
almost a $70\%$ fraction of the sky when $\rho_{max}=15$. For
$\rho_{max}\simeq 30$ the  detection is likely almost in any direction.  

\begin{figure}
 \includegraphics{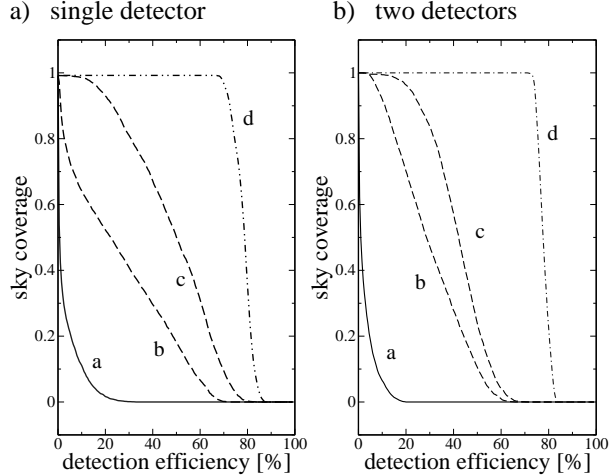}
\caption{\label{2sphereskycoverage}   Fraction  of  the   sky   as  a
  function of the detection efficiency for a GW  burst of deposited energy
$T_{GB}= 0.6\mu K$ and $\rho_{max}=6$ for a single spherical antenna {\bf i)} and two optimally
oriented spherical  detectors {\bf ii)} located at  Leiden and
Sa$\tilde{o}$ Paulo. Curve  {\bf a} corresponds  to a
read-out configuration  with a single  transducer located in position 1 of each detector, curves  {\bf b}
and {\bf  c} to  a three
transducers    configuration   with   $\theta_{TI}=37.3773^{\circ}$   and
$\theta_{TI}=73.1876 ^{\circ}$  and curve {\bf  d} to a  complete spherical
detector with  6 transducers.}
\end{figure}

\Fref{2sphereskycoverage} presents the fraction of sky as a function
of the detection efficiency of a GW  burst of deposited energy
$T_{GB}= 0.6\mu K$ for different transducer configurations.
Plot {\bf i)} refeers to a   single spherical antenna, while plot {\bf
  ii)} shows the detection  efficiency of  two optimally
oriented spherical  detectors located at  Leiden and
Sa$\tilde{o}$ Paulo.  This  was  obtained  by  maximizing the  portion  of  sky
simultaneously seen by both detectors as a function of the orientation
of  the single  detector reference  system with  respect to  the local
south. This  corresponds to a  rotation of the brasilian  detector lab
frame $(x,y,z)$ (see \fref{sphere_TI})  around the local z-axis  of about
$\phi_0=-135^{\circ}$.

A detector with a single trasducer is unable to detect any signal from
the sky when the SNR is as low as $\rho_{max}=6$. For the same SNR,  the sky coverage of
a single detector with three and six transducer is respectively more than
$40\%$ and $100\%$ with $50\%$ detection efficiency. This is mainly
due to the increase of detector bandwidth when more than one
transducers are used. \Fref{2sphereskycoverage}.ii shows that for detector with one or three transducers, the
twofold coincidence probability remains lower than the detection
efficiency of a single sphere. Two spherical antenna  with six
transducers  can equally detect  any sources in the sky even when the
SNR is as low as $\rho_{max}=6$. 

\subsection{Sensitivity of a sphere with not ideal transducers}

It has been  shown that for a perfect  sphere the resonators mistuning
and misplacing  has little effect  on the isotropy when  the deviation
from the ideal TI  configuration is less than $1\%$ \cite{Stevenson98,
MerkLobo99}.   We study  here two  other possible  degradation effects
which may arise in a real  spherical detector. The first is related to
the broken spheroidal  mode degeneracy and the second  to the the fact
that  a  real  transducer  is   not  a  pointlike  mass  as  generally
considered.   In a  real detector  the spheroidal  mode  degeneracy is
broken due  to the  suspension and  the holes made  on its  surface to
house the transducers. The modes spreading can be as large as $5\%$ of
the main  resonance.  It  becomes natural to  ask to which  modes each
transducer should be tuned and how  good the tuning should be to avoid
a sensitivity  degradation.  We consider here a  numerical analysis of
the transducer  mechanical and electrical mistuning for  a sphere with
six resonators  with a mass  ratio $\mu=0.01$ and operating  at nearly
the quantum limit.

\begin{figure}
 \includegraphics{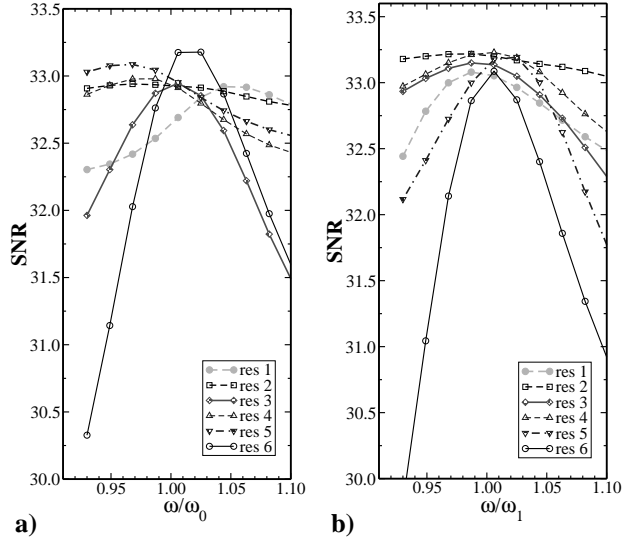}
 \caption{\label{SNRresfreq}  Resonators mistuning.   The  graph shows
the detector SNR  for a $T_{GB}=3\mu K$ one-cycle circularly
polarized GW burst
as a  function of the  mistuning parameter $\omega/\omega_0$  for each
resonator in the case of {\bf  a)} a starting arbitrary set of natural
resonances $\omega_0=2\pi[987, 1001, 1001, 1008, 1012, 1017]$ and {\bf
b)} an  optimized set $\omega_1=2\pi[1047, 971, 980,  970, 960, 1027]$
obtained after few tuning iterations.}
\end{figure}

In  \fref{SNRresfreq}a),  we   see  the  effect  on  the   SNR  for  a
$T_{GB}=3\mu  K$  circularly  polarized  GW  burst  when  the  natural
resonances  of each  resonator are  modified from  the  initial values
$\omega_0=2\pi[987, 1001, 1001, 1008, 1012, 1017]$, arbitrarily chosen
equal to each  of the spheroidal mode resonances.  One finds a maximum
change  of  $10\%$ in  the  SNR for  a  resonator  mistuning of  about
$10\%$.  It  is  possible  to  optimize the  tuning  by  shifting  the
resonator resonance frequency  as much as indicated by  the maximum of
the  $SNR$  in \fref{SNRresfreq}.a.  This  procedure  can be  repeated
several times.  The result is shown in  \fref{SNRresfreq}.b, where the
SNR is maximum for each resonator around the new set of the resonators
natural   frequencies  $\omega_1=2\pi[1047,   971,   980,  970,   960,
1027]$. Such  an optimal frequency  set derives from a  combination of
multiple coupling, modes splitting  and transducers position. Once the
bare sphere  spheroidal modes and  the resonator masses  and positions
are known, one  can always find an optimal  natural resonance for each
resonator.

We remark  here that,  due to the  presence of  so many modes  and the
related  multiple splitting,  it  could be  difficult  in practice  to
determine the resonators frequency with an accuracy better than $10\%$
and one has  probably to accept a not optimized  detector. The loss in
sensitivity is in  any case less than $10\%$ for  a $100 Hz$ mistuning
of the mechanical modes of a detector resonating at $1 kHz$ and with a
tuned matching network.

So  far  we consider  the  transducers  operating  on the  sphere  all
identical except for their  main resonance frequency. Here we evaluate
the  effect  on  the   detector  anisotropy  in  the  sensitivity  and
bandwidth, which  derives from using six not  identical transducers to
read-out  the  five  quadrupolar  mode  of the  sphere.  We  arbitrary
modified the  parameters of each  transducer such as  mass, mechanical
and  electrical  quality  factor, transducer  capacitance,  electrical
coupling  factor and SQUID  noise of  a maximum  of $15\%$  from their
optimal value.  As shown in  \fref{6trSNR}.b, the detector SNR,  for a
circularly polarized  GW burst with $T_{GB}=3\mu K$,  is reduced about
the $30\%$.  This is mainly due  to a decrease in  signal bandwidth as
one can see from \fref{6trBW}.b.

In the following we study  how the detector sensitivity decreases when
the transducer  electrical resonator is  not perfectly matched  to the
mechanical resonator.  Such a situation  could arise in  practice when
the  electrical mode  cannot be  arbitrarily adjusted  to  the optimum
value define by \eref{conditions} as  a consequence, for example, of a
voltage  leakage  in  the  bias  lines.   One  finds  that,  when  the
electrical mode of  only one transducer is mistuned,  even up to about
$30\%$ of the optimal frequency, very little effect is observed in the
SNR and bandwidth.
\begin{figure}
 \includegraphics{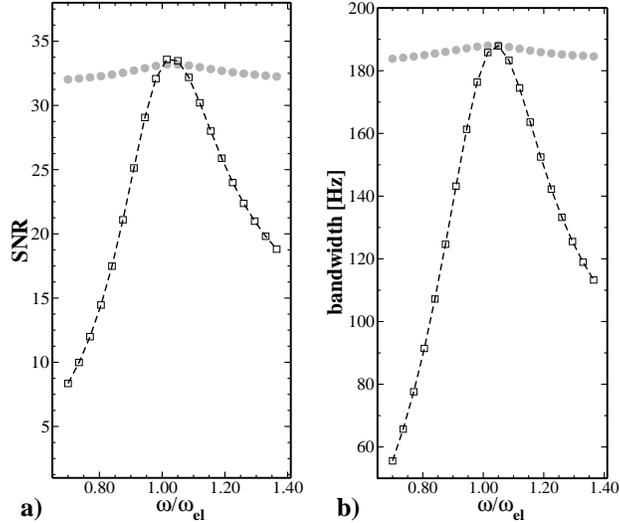}
 \caption{\label{Eltuning}  Mistuning  of  the electrical  mode.   The
graph shows the detector SNR  (a) and bandwidth (b) for a $T_{GB}=3\mu
K$  one-cycle circularly polarized  GW  burst  as a  function  of the  
mistuning parameter  $\omega/\omega_{el}$ where  $\omega_{el}$ is  equal  to the
optimized mechanical resonator frequency set. The data show the effect
when  the electrical  mode of  only  one transducer  is detuned  (full
circles) and when the electrical  modes of all the six transducers are
equally mistuned (open squared).}
\end{figure}   
This  is  evident   from  the   full  circle   data  of
\fref{Eltuning},  where   the  detector   SNR  and  bandwidth   for  a
$T_{GB}=3\mu K$ circularly  polarized GW burst is given  as a function
of the  mistuning parameter $\omega/\omega_{el}$,  being $\omega_{el}$
equal  to the  optimized  mechanical resonator  frequency set  derived
above. When all the electrical modes are decoupled from the mechanical
modes, the SNR and bandwidth  decrease of a factor proportional to the
mistuning factor as shown by the open squared data of \fref{Eltuning}.

In the  numerical analysis of resonant detectors  one always considers
the mechanical  resonators as point-like masses. A  real transducer is
sampling  the  sphere  surface  radial  displacement  in  many  points
belonging to the contact surface  between the resonator spring and the
sphere.   The actual displacement  can be  seen as  the results  of an
average  of  such  a   sampling.   We  calculated  here  the  detector
sensitivity  when a  real  resonator is  considered.   Three kinds  of
resonators  are  generally   used  in  resonant  detectors:  mushroom,
membrane and "rosette" resonators. All of them have in common the fact
that the spring  is attached to a support,  which is rigidly connected
to  the antenna's surface.   The variable  read-out considered  in our
calculation,  in  analogy  with  \cite{Levin98},  can  be  written  as
\beq{Xreadout}                                                
x(t)=\int
f(\mathbf{r_t,r_r})y(\mathbf{r_t,r_r},t)dr_tdr_r       
\eeq      
where
$y(\mathbf{r},t)$  is  the  radial  displacement of  the  intersection
points between the support and the springs of the mechanical resonator
used to  amplify the sphere  displacement.  $f(\mathbf{r})$ is  a form
factor   and  the   integral   is  calculated   along  the   resonator
support-spring intersection.  We consider transducers with cylindrical
symmetry where $r_t$ is  the radius of the support-spring intersection
and  $r_r$  gives  the  radial  position of  the  intersection  points
referred  to  the  sphere   centre.   For  "mushroom",  "rosette"  and
"membrane"  resonators  those   intersection  are,  respectively,  the
"mushroom" leg section, the  cylindrical section where the membrane is
attached  to the  resonator  ring support,  and  the "rosette"  spring
sections at the attachment point with the ring support.

The form of the radial  displacement depends strongly on the resonator
geometry  and  springs  topology,  making  it  difficult  to  find  an
analytical  expression  like  the  one  for the  gaussian  laser  beam
considered  in \cite{Levin98},  and \cite{Brian03}.   In the  case when
only the first spheroidal modes  are considered, the form factor for a
"membrane"  and  "rosette"  transducer  becomes  \beq{formfactor_memb}
f_{memb}(\mathbf{r_t,r_r})=\frac{s}{2\pi                      r_t\delta
r_r}\frac{A_{12}(r_r)}{A_{12}(r_s)},  \eeq where  $\delta r_r$  is the
spring thickness  and $A_{12}(r)$ is the  spheroidal quadrupole radial
amplitude  function  described  in  \cite{Lobo95}.  For  a  "membrane"
transducer    $s=1$,     while    for    a     "rosette"    transducer
$s=S_{ros}/S_{memb}<1$ is the  ratio between the intersection surfaces
defined by the "rosette" springs and a membrane of the same thickness.

In the case of  a "mushroom" transducer, we have \beq{formfactor_mush}
f_{mush}=\frac{1}{\pi  r_t^2}\frac{A_{12}(r_r)}{A_{12}(r_s)}.  \eeq In
proximity  of  the  sphere  surface  the radial  amplitude  is  slowly
changing  and  the ratio  $A_{12}(r_r)/A_{12}(r_s)$  can generally  be
approximated to 1.
 
\begin{figure}
 \includegraphics{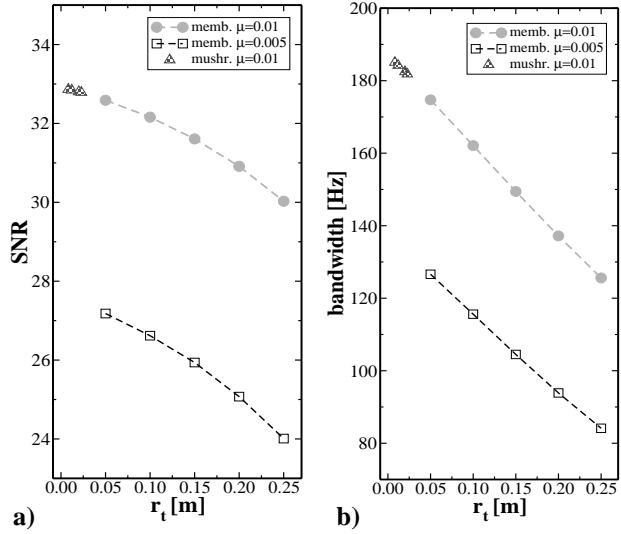}
 \caption{\label{SNRBWrealRes} Detector $SNR$  and signal bandwidth as
a function  of the resonator  radius $r_t$ for a  "membrane" resonator
with  mass  ratio  $\mu=0.005$,  open squares,  and  $\mu=0.01$,  full
circles, and a "mushroom" resonator with $\mu=0.01$, open triangles. }
\end{figure}

\begin{figure}
 \includegraphics[scale=0.95]{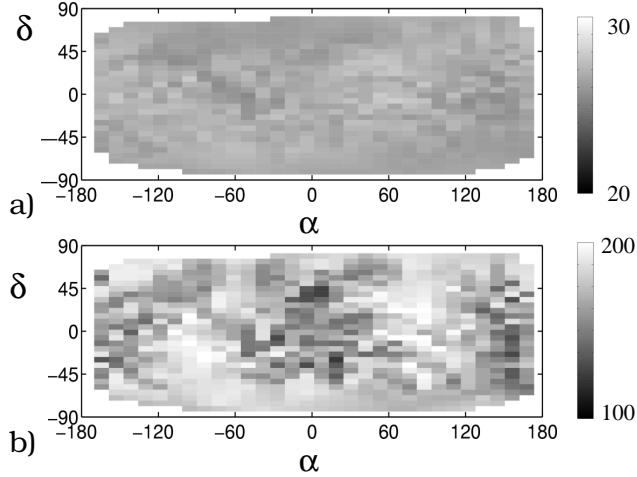}
 \caption{\label{SNRBWanisotropy} Detector  $SNR$ and signal bandwidth
anisotropy for a circular polarized GW burst with $T_{GB}=3\mu K$ when
a "rosette" resonator with $r_t=0.2 m$  is used to read out the sphere
quadrupolar modes.}
\end{figure}

 In \fref{SNRBWrealRes} the SNR and bandwidth for a circular polarized
$T_{GB}=3\mu K$ GW  burst of a nearly quantum  limited sphere with six
resonators is shown  as a function of the  resonator radius $r_t$. The
improvement  in sensitivity,  obtainable as  described above  by using
massive  resonator, with  $\mu =  0.01$,  is slightly  reduced by  the
resonators large  radius $r_t\sim 0.2  m$. The SNR decreases  of about
$10 \%$ in  this case. A "mushroom" resonator  is preferable, as shown
in \fref{SNRBWrealRes}, but high Q massive resonators are difficult to
achieve for such a geometry.
 In the sky maps shown in \fref{SNRBWanisotropy}, obtained
from  1000  randomly distributed  events  of  circularly polarized  GW
bursts,  the  SNR  and  bandwidth  anisotropy for  a  transducer  with
$r_t=0.2 \, m$ is  shown. One finds up to $15 \%$  of asymmetry in the
SNR and up to $40 \%$ in the detector bandwidth.

\section{Calibration of a spherical gravitational wave detector}
\label{calibration}  To calibrate  a gravitational  wave  detector one
needs to postulate  a model of the complete  detector whose parameters
are experimentally  determined.  The  calibration is performed  in two
steps. First the effective temperature of the modes is estimated. This
is  important in  order to  understand  the detector  dynamics and  to
estimate the  noise contributions.  Secondly,  one has to  measure the
detector  response  to  an  applied  force,  which  is  equivalent  to
evaluating    the    transfer    function   $\mathbf{G_{sig,I}}$    in
\eref{strainnoise}.    We   stress    here   that   a   full   antenna
characterization  implies the estimation  of the  terms of  the matrix
$\mathbf{G_{sig,I}}$, which relates the input signal and noise to each
transducer  SQUID input  current. It  can be  achieved by  injecting a
known signal at each step  of the transducer chain, i.e: radial forces
to  the five spheroidal  modes of  the sphere,  radial forces  to each
resonator,  voltage  signal  in  the superconducting  transformer  and
voltage  signal  at  the  SQUID  input.   In  order  to  evaluate  the
sensitivity of the antenna to GW signal, however, one needs especially
to measure the transfer function  which converts the forces applied to
the spheroidal modes to the SQUID input current of each transducer. By
measuring the  noise of each  transducer during normal  operation, one
can finally estimate the strain sensitivity as in \eref{strainnoise} .

We describe below  two methods that can be used  to perform a complete
calibration of  a spherical detector  in analogy with the  method used
for the bar detector AURIGA \cite{Baggio02,Baggio05}
  
\subsection{Modes equivalent temperature}

According  to the  fluctuation  dissipation theorem,  the voltage  and
current  power spectra observed  in an  electrical circuit,  when only
thermal noise  sources are presents, are given  by \beq{FDTVI} S_V=4kT
Re(Z(\omega)), \quad  S_I=4kT Re\left(\frac{1}{Z(\omega)}\right), \eeq
where $Z(\omega)$ is the impedance  seen at the input of the amplifier
used  for the  read-out and  T is  the thermodynamic  temperature.  By
measuring the  transducer output spectrum in normal  operation and the
input  impedance  of  the  circuit  one can  evaluate  the  equivalent
temperature of the transducer chain. If only thermal noise is present,
the  equivalent  temperature of  the  chain  should  be equal  to  the
thermodynamic temperature of the experiment.  Because the SQUID is not
an ideal amplifier, its current  and voltage noise give a contribution
to the total transducer output  noise. This contribution is in general
not negligible. The total current  noise measured at the input of each
transducer line  SQUID can be written  as \cite{Baggio05}, 
\beq{SItotZ}
S_I=    4kT\frac{Q_a}{Q_{el}}Re   \left(\frac{1}{Z(\omega)}\right)   +
\frac{S_{vv}(T)}{|Z(\omega)|^2}+S_{Iii}(T), 
\eeq 
where the second term
on  the right  hand side  of  the equation  is the  back action  noise
contribution  from   the  SQUID  amplifier  with   $S_{vv}$  given  by
\eref{CGT}.   $Q_a$ is  the apparent  quality factor  produced  by the
damping and $Q_{el}$ is the intrinsic quality factor of the electrical
matching network. The factor $\frac{Q_a}{Q_{el}}$ in the thermal noise
appears  when  the cold-damping  network  is  active  in the  read-out
circuit \cite{Baggio05}.   It comes  of the fact  that the  damping is
only a lossless electronic feedback effect and there is no dissipation
associated to it.  From \eref{SItotZ} we see that when the SQUID noise
parameters $S_{vv}(T)$,  and $S_{ii}(T)$ are known,  and the impedance
$Z(\omega)$ is measured, from the  fit of the SQUID output current one
can estimate as a fitting parameter the equivalent temperature of each
transduction chain.   One can measure the  input impedance $Z(\omega)$
seen by  the SQUID by injecting a sinewave signal with
defined level through  a calibration coil weakly coupled  to the SQUID
input  circuit.

We can  simulate numerically  the calibration procedure.   Denoting by
$M_{cal}$ the  mutual inductance between the calibration  coil and the
SQUID input circuit and by $R_{cal}$ the resistance of the calibration
line,  to estimate  the impedance  $Z(\omega)$  one has  to solve  the
system  of equations in  \eref{eqn_of_motion}.  All  the terms  on the
right hand side  are zero but the SQUID input  voltage, which is given
by  $V_n=\frac{j\omega M_{cal}}{R_{cal}}V_{cal}$,  with  $V_{cal}$ the
voltage of  the injected calibration signal.  By  measuring the output
response of the SQUID amplifier  we obtain a direct estimation of each
transducer admittance from the following weighted average: 
\beq{Zmeas}
\frac{1}{Z^{nn}(\omega)}=
\frac{Re\{I_nV^{*}_n\}+jIm\{I_nV^{*}_n\}}{|V_n|^2}=G^{nn}_{V_n},   
\eeq
where  $I_n$  is   the  current  at  the  input   of  each  SQUID  and
$\mathbf{G_V^{nn}}$  was  defined  in \eref{Isqinput}.   The  weighted
average gives  a more precise result than  simply measuring $I_n/V_n$.
We remark that both amplitude and phase of the input and output signal
have to be measured to estimate $Z(\omega)$.

The  admittance  $1/Z(\omega)$  can   be  approximated  by  a  complex
polynomial expansion  with $N_p=5+2\times N$  poles and $N_q=5+2\times
N-1$          zeros          as         follows          
\beq{polyexp}
\frac{1}{Z(\omega)}=A\frac{\Pi_{k=1}^{N_q}
(j\omega-q_k)(j\omega-q_k^*)}{\Pi_{k=1}^{N_p}(j\omega-p_k)
(j\omega-p_k^*)}(j\omega)^{(N_p-N_q)}.    
\eeq   
If   $\omega_k$   and
$Q_{a,k}$ are the resonant frequency and the apparent Q-factor of each
measured      resonance      oft      he     system,      we      have
$p_k=j\omega_k-\omega_k/2Q_{a,k}$.The  zeros $q_k$  can be  written in
the same  form and  have frequency  and Q which  depends on  the modes
coupling.  Once  the admittance is  measured, a polynomial fit  can be
performed in order to find its zero and poles.

We simulated  numerically the mode  temperature calibration procedure.
The current  noise at the input  of the SQUID  amplifier was estimated
from \eref{FDTVI}  by calculating the  impedance seen by the  SQUID in
the  form of  the  admittance $1/Z(\omega)$.  For  simplicity we  only
consider two resonators located in position 1 and position 2 of the TI
arrangement. The  detector is at  $T= 100 \,mK$ with  a $T/Q=2.5\times
10^{-7}$, a  SQUID energy  resolution of $200  \hbar$ and  an apparent
quality factor $Q_a=600$. The  electrical modes of each transducer are
tuned to the mechanical modes.

\begin{figure}
 \includegraphics[scale=0.95]{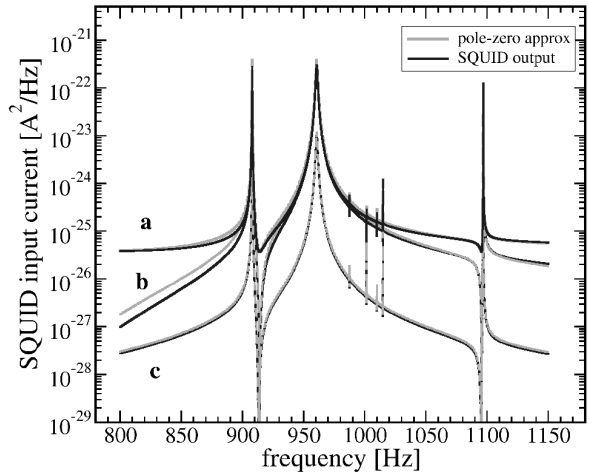}
 \caption{\label{SisqZ}  Mode  temperature  calibration.  The  current
noise  density at  the input  of transducer  1 SQUID  is  estimated by
calculation the admittance matrix $1/Z(\omega)$. The contribution from
thermal noise  {\bf b)},  back-action noise {\bf  c)} and  total noise
including the  additive SQUID  current noise {\bf  a)} are  shown. The
dashed curves are obtained from the pole-zero polynomial expansion.}
\end{figure}

\Fref{SisqZ} shows the SQUID input current noise contribution from the
thermal  and back-action  noise and  SQUID additive  current  noise of
transducer line 1.
  The zero-pole approximation  of \eref{polyexp}  is shown
with   dashed  curves   on  top   of  the   simulated   current  noise
contributions.
 
\subsection{Force calibration and transfer functions measurements}

To calibrate a spherical detector one should be able to experimentally
evaluate the  transfer functions which relate the  output current with
the GW force acting on the  spheroidal modes.  When a force is applied
at a point  $(\theta_c,\phi_c)$ on the surface of  the sphere, all the
$5+2N$ modes of the sphere and transducers are excited at an amplitude
which depends on the calibrator position, the resonators positions and
the coupling  between resonators and spheroidal  modes.  A calibrator,
consisting  of  a  capacitive   transducer  with  the  main  resonance
frequency  $\omega_{cal}$ not tuned  with the  fundamental frequencies
$\omega_{0}$ of  the detector,  can be used  to convert  an electrical
signal  into  a  constant  force  acting on  the  sphere  modes.   For
$\omega_{cal}>>\omega_{0}$,  such  a  force  is  given  by  
\beq{Fcal}
F_c(\omega)=\frac{C_{cal}E_{cal}}
{1-\frac{C_{cal}E_0^2}{m_{cal}{\omega_{cal}^2}}}V_{cal}(\omega),    
\eeq
where  $C_{cal}$,   $E_{cal}$  and  $m_{cal}$   are  respectively  the
capacitance,  the  dc  bias  electric   field  and  the  mass  of  the
calibrator. $V_{cal}(\omega)$  is the excitation voltage  given to the
calibrator at the resonance frequencies of the detector.

A force acting  on the sphere surface always  excites a combination of
the 5  spheroidal modes. If denote  by $\mathbf{F_c}=(F_1...F_{Nc})$ a
vector   of  $N_c$  radial   forces  applied   to  the   $N_c$  points
$P_c=(\theta_c,\phi_c)$,  with $c=1...N_c$,  the  corresponding forces
$\mathbf{F_m}$ acting  on the 5  spheroidal modes can be  described by
the    calibrators   pattern    matrix    $\mathbf{B_C}$:   \beq{FcFm}
\mathbf{F_m}=\alpha \mathbf{B_C}\mathbf{F_c}.   \eeq From \eref{BmjYm}
we  find ${B_C}_{m,c}=Y_{m,c}(\theta_c,\phi_c)$.   The  pattern matrix
$\mathbf{B_C}$   has  the   same  physical   meaning  as   the  matrix
$\mathbf{B}$,  but  is referred  to  the  calibrator positions.   From
\eref{FcFm} it is  clear that the gravitational forces  acting on each
of the five quadrupole modes,  distributed over the entire volume, can
be simulated  by a linear combination  of radial forces  acting on the
$N_c\geqslant  5$  calibrators.   This  conclusion  is  based  on  the
assumption that  the real  spherical antenna dynamics  is the  same as
that of an ideal sphere that can fully be modelled using the spherical
harmonics approach  described above.  The validity  of this assumption
could  be evaluated  through  a  finite element  analysis  (FEA) of  the
detector  structural  model or  by  experimentally  measuring the  six
transducers  response  to several  linear  combinations of  excitation
given through the calibrators.
 
To operate  the sphere  as a GW  detector, only one  force calibrator,
mounted at an  arbitrary position on the sphere  surface, is necessary
when $N=6$ transducers are used in  the TI configuration. This is due to
the existing  one-to-one relation between the {\it  mode channels} and
the  forces acting  on  the  spheroidal modes.   It  is convenient  to
proceed by transforming the  $N$ measured transducers outputs into the
$5$  {\it mode  channels}, which  directly describe  the  GW amplitude
acting on the spheroidal modes \cite{Merkowitz95}. In such a framework
we derive the optimal filter for  each {\it mode channel} to define an
operative procedure  to signal detection with a  GW spherical resonant
antenna.  In order  to do this we define the  $5\times 5$ mode channel
noise       spectral      density       matrix       as      
\beq{SBI}
\mathbf{S_{I,B}}(\omega)=\mathbf{B}\mathbf{S_I}
(\omega)\mathbf{B^\dagger},
\eeq 
and, using  the admittance matrix $\mathbf{G_{sig,I}}$ introduced
in  \eref{strainnoise}, the  $5\times  1$ mode  channel signal  vector
\beq{modechannel_sig} 
\mathbf{I_{g,sig}}=\mathbf{B}\mathbf{I_i}=\alpha
\mathbf{B}\mathbf{G_{sig,I}}\mathbf{B_C}\mathbf{F_c}.      
\eeq     
By applying  a known  constant  force to  the calibrator  $\mathbf{F_c}$,
using \eref{FcFm}  we get the  linear combination of  spheroidal modes
forces generated by the calibration force. By measuring simultaneously
amplitude and phase  of the six transducer outputs  we obtain the {\it
mode  channel}   response  from  \eref{modechannel_sig},   and,  using
\eref{strainnoise},  estimate the  detector  strain sensitivity.   The
matrix  $\mathbf{S_{I,B}}(\omega)$ in  \eref{SBI} is  diagonal because
the  {\it  mode channels}  are  statistically  independent.  The  only
assumption  made here  is  that the  the  pattern matrix  $\mathbf{B}$
derived above for an ideal sphere can be used to describe the dynamics
of the real  detector with six transducer in  the TI arrangement. This
can  be experimentally  verified in  a separate  cryogenic experiment,
using for  example a  set of six  calibrators in the  TI configuration
with  a $60^{\circ}$ de-phase  of the  azimuthal angle  with respect  to the
transducer arrangement.  Each spheroidal  mode force can be reproduced
by simultaneously applying a  linear combination of constant forces on
the  six  calibrators,  as   described  by  \eref{FcFm}.   
When  $N<5$
transducers  are used we  cannot transform  the transducer  outputs in
mode channels.  A set of at  least 5 calibrators is  then necessary to
experimentally   measure  the   transducer  response   to   each  mode
channel. If only one calibrator  is used, the detector output can only
be  calibrated for a  particular combination  of the  five quadrupolar
forces     $F_m$.     

Each    element     on    the     diagonal    of
$S^{(m)}_{I,B}=I_m(\omega)I^\ast_m(\omega)$ can be written in terms of
polynomial ratio where  the poles are the same as  the ones derived in
the impedance measurement  of each transducers,see \eref{polyexp}, and
the  zeros depend  on  the current  function  $I_m(\omega)$. From  the
factorization    of    $I_m(\omega)$    one    finds    
\beq{polyexpI}
I_m(\omega)=S^{1/2}_{0,m}\Pi_{k=1}^{N_p}\frac{
(j\omega-q_{k,m})(j\omega-q^\ast_{k,m})}{(j\omega-p_k)(j\omega-p^\ast_k)},
\eeq 
where $S_{0,m}$  is the wideband noise of  the $m^{th}$ {\it mode
channel}, and  equals the  amplifier additive  white noise  if  all the
transducers   SQUIDs   are   identical.   From   the   definition   in
\eref{polyexpI},  $S^{(m)}_{I,B}$ is  real for  real $\omega$  and the
number of  zeros and poles  is the same  as a consequence  of assuming
$S_0^m$to be purely white.

The transfer functions for a GW signal, which convert the quadrupolar modes 
forces into {\it mode channels} currents according to \eref{modechannel_sig}, 
contain the same poles $\{p_k\}$ and their factorization becomes
\beq{polyexpHm}
H_m(\omega)=H_{m,cal}(\omega)\frac{\Pi_{k=1}^{N_r} 
(j\omega-r_{k,m})(j\omega-r^\ast_{k,m})}{\Pi_{k=1}^{N_p}
(j\omega-p_k)(j\omega-p^\ast_k)}.
\eeq   
In the equation above  $N_p>N_r$ and $H_{m,cal}(\omega)$  is a force 
calibration constant which has to be experimentally determined at 
each cool  down.

From now on we apply   to each {\it mode channel} the standard 
Wiener-Kolmogorov (WK) filtering operation, developed so far for bar 
detectors \cite{PallPizz81}. We follow here the approach described in 
\cite{Baggio02}. 
The best linear estimate of the amplitude $h_0$ of a given signal 
$h(t)$, with max ${h(t)}=h_0$ at the arrival time $t=0$, 
buried into an additive, zero mean, stationary gaussian noise $\eta$ can be 
obtained by correlating the mode channel output 
$Y_m(\omega)=\tilde{h}(\omega)H_m(\omega)+\eta (\omega)$  to the matched 
WK filter \cite{vantrees},
\beq{WKfilter}
W_m(\omega)=\sigma^2_A\frac{H^\ast_m(\omega)\tilde{h}^\ast(\omega)}
{S^{(m)}_{I,B}(\omega)},
\eeq  
where $\sigma^2_A=\int d \omega|H_m(\omega)
\tilde{h}(\omega)|^2/S^{(m)}_{I,B}(\omega)$ is the variance of the noise 
after the filtering.
The WK filter  splits as the product $L_m(\omega)M_m(\omega)
\tilde{h}^\ast(\omega)$, where $L_m(\omega)$ is a whitening filter  
for the noise $S^{(m)}_{I,B}(\omega)$ given by
\beq{Lm}
L_m(\omega)=I^{-1}_m(\omega)=S^{-1/2}_{0,m}\Pi_{k=1}^{N_p}
\frac{(j\omega-p_{k,m})(j\omega-p^\ast_{k,m})}{(j\omega-q_{k,m})
(j\omega-q^\ast_{k,m})}.
\eeq
$M_m$ is a bandpass filter around the frequencies $\omega_k=|Im(q_{k,m})|$ and 
 bandwidths $\Delta \omega_k=2Re(q_{k,m})$. Such a bandwidth is generally much 
larger than the intrinsic bandwidth of each resonance and, in the case of a 
transducer with coupled electrical modes, it can reach values as large as 
$\sim 200 Hz$.  One finds
\beq{Mm}
M_m(\omega)=\sigma^{2}_A\cdot S^{-1/2}_{0,m}\frac{\Pi_{k=1}^{N_r}
(j\omega+r_{k,m})(j\omega+r^\ast_{k,m})}{\Pi_{k=1}^{N_p}(j\omega+q_{k,m})
(j\omega+q^\ast_{k,m})}.
\eeq
The product $L_m(\omega)M_m(\omega)$ is the WK filter for a delta-like GW 
pulse. The extra term $\tilde{h}(\omega)$ should be added when a general GW 
signal $h(t)$ is considered.
It can be shown that for a resonant bar detector, $N_r=1$ and $r_1=0$. In this 
case the WK filtering procedure is then fully defined by the zeros  ${q_k}$ and
 poles ${p_k}$ of the noise power spectrum $S_I(\omega)$, the additive 
amplifier white noise $S_0$ and the calibration constant $H_0(\omega)$. 
For a spherical detector this is true only if the quadrupolar modes degenerate 
into a single resonant frequency and all the transducers have the same 
resonance. When multiple resonances are present, as  is the case for real 
spherical detectors, the zeros in \eref{polyexpHm} do not cancel i.e., 
$N_r>1$ and $r_{k,m}\neq 0$.
 This is due to the fact that when the quadrupolar modes are non-degenerate, 
a mixing occurs between {\it mode channels}. A fraction of the signal which 
should only go to one {\it mode channel} leaks into the others. 
The WK filtering procedure should include the extra set of parameters 
$r_{k,m}$, whose total number depends on the mode and transducer considered 
and must be experimentally determined.

\begin{figure}
 \includegraphics{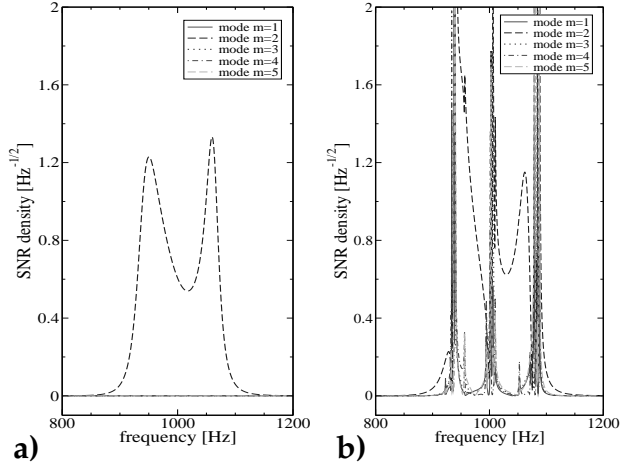}
 \caption{\label{SNR_5modes} {\it Modes channels} response from a simulated  
linearly polarized burst coming along the $z$ direction in the detector frame 
for a  detector with degenerate,  {\bf a)}, and non-degenerate, 
{\bf b)}, modes. For a detector with degenerate modes, {\bf a)}, only the 
second  mode gives a  larger signal than the noise. For a real detector 
with non-degenerate modes{\bf b}, a mixing between the modes is 
present around the resonances making the analysis more complex }   
\end{figure}

\begin{figure}
 \includegraphics[scale=1.1]{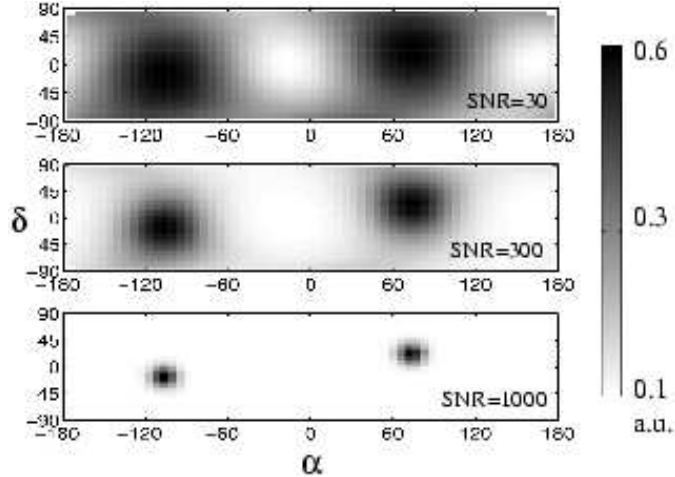}
 \caption{\label{wavedirection} Signal direction reconstruction. 
An overall sky search is performed calculating the likelihood 
function $\lambda$.  We applied simulated burst signals with
$SNR\sim 30,\, 300,\, 1000$, amplitude $h_0=0.34\cdot
10^{-20},\,1.1 \cdot
10^{-20},\, 2\cdot
10^{-20}$ and $\tau_{GB}\sim 1 ms$,  
linearly polarized and coming from a source at declination 
$\delta=20^{\circ}$ and ascension $\alpha=70^{\circ}$. A single 
spherical detector cannot distinguish between sources laying in 
two opposite  hemispheres.}
\end{figure}

In \fref{SNR_5modes} the mode channels response is plotted  for a linearly 
polarized burst coming along the detector $z$ direction. 
For a detector with degenerate modes, \fref{SNR_5modes}.a, only the second 
mode gives a  larger signal than the noise, as can be expected for a signal 
from that particular direction and polarization. For a realistic detector 
with non-degenerate spheroidal modes a mixing between the modes is 
present around the resonances making the analysis more complex, see 
\fref{SNR_5modes}.b). However, the energy stored in the second mode, 
which can be derived from the integral of the SNR density, is larger 
than in the others and the incoming wave direction can still be reconstructed 
without any significant accuracy loss with respect to the degenerate case.
 To estimate the incoming wave direction we can use the approach  
derived in \cite{ZhouMich95}, using standard theory of signal detection. 
After being optimally filtered, the 5  mode channels generate a set of 5 
amplitudes,  $g_m$ , for an input GW burst. The likelihood function 
for a detector with stationary and Gaussian noise is given by
\beq{LSerror}
\lambda=\frac{1}{2\pi}\prod^5_{m=1}exp\left(-\frac{\left[h_m-g_m\right]^2}
{\sigma_m^2}\right), 
\eeq
where $h_m$ is the expected gravitational wave signal amplitude, 
$g_m$ is the mode channels amplitude obtained after the WK filtering 
procedure described above and $\sigma_m$ is the variance of $g_m$. 
A likelihood map is generated by plotting this function in the 
declination-ascension plane, $(\delta,\alpha)$. The maximum value 
of $\lambda$ gives to the estimated wave direction.
In the example below, a realistic detector at the quantum limit,  
with optimized parameter is considered.
\noindent We applied simulated burst signals, $1\, ms$ long and with amplitude $h_0=0.34\cdot
10^{-20},\,1.1 \cdot
10^{-20},\, 2\cdot
10^{-20} $ and  $SNR\sim 30,\, 300,\, 1000$, 
with linear polarization $h_+=h_0$ and $h_\times=0$, coming from a source at 
declination $\delta=20^{\circ}$ and ascension $\alpha=70^{\circ}$. 
\Fref{wavedirection} plots the resulting likelihood functions.
 
The accuracy in the direction and polarization estimates depends on the 
SNR and is equal to $\Delta \Omega=2\pi/SNR$ \cite{ZhouMich95}, \cite{Stevenson98}; it is direction independent 
when a sphere with more than 5 identical, point-like transducers is considered 
\cite{MerkLobo99}.  As it comes clear from the plot, a single spherical 
detector cannot distinguish between sources laying in the two opposite 
hemispheres.
In the example described above we assumed for simplicity that the signal 
polarization were known. When this is not the case, one should include 
also the first Euler angle $\psi$ as a variable to estimate the 
likelihood ratio. 

As discussed  in \cite{Stevenson98}, for  a known signal,  the optimal
detection strategy  for a {\it  vector} output detector like  a sphere
with  N transducers  is to  compute the  value of  the  optimal linear
filter with  {\it scalar} output  $\nu$ which maximizes the  SNR. This
procedure is  useful if one wants  to monitor the total  energy in the
sphere and claim a detection only on the base of an excess of absorbed
energy. No direction information is possible in this way.
 
If we let $\mathbf{I}={I_1...I_N}$ be the transducers output stream, a
linear filtering  operation is  performed which, in  frequency domain,
can        be        described        by        
\beq{filtering_vector}
\nu(\omega)=\mathbf{W^\dagger}(\omega)\mathbf{I}(\omega),  
\eeq  
where
$\mathbf{W}$ is a vector transfer function which maximizes the SNR and
was      found     \cite{Stevenson98},     to      be     
\beq{Wvector}
\mathbf{W}(\omega)=\mathbf{I^\dagger_{sig}}(\omega)
\mathbf{S^{-1}_I}(\omega).
\eeq
In  the  above,
$\mathbf{I_{sig}}=\mathbf{G_{sig}}(\omega)\mathbf{F^S_m}$    is    the
current generated in each transducer  by a force $F^S_m$ on the sphere
generated by a  GW signal. As shown in  \cite{ZhouMich95}, the maximum
SNR for a multichannel detector is  the sum of the maximum SNR of each
individual  channel. The  optimal linear  filter introduced  above for
each {\it mode channel} is then  useful in order to compute the scalar
output $\nu$ which maximizes the SNR.
\section{Conclusions}
\label{conclusions}

We  derived a  complete and  detailed electro-mechanical  model  for a
spherical  gravitational   wave  detector  operating   with  multiple,
two-mode capacitive transducers where  the electrical resonant mode of
a superconducting matching  LC resonator can be tuned  to the resonant
modes.  The  signal current form the  matching network is  read out by
SQUID  amplifiers.   The model  allows  to  numerically calculate  the
sensitivity of a realistic detector and to study in detail the effects
of the  main mechanical and electrical parameters  of the displacement
read-out system on the strain sensitivity and bandwidth. All the known
noise sources are discussed and considered in the model.

A  complete numerical analysis  has been  performed for  a 2  meter in
diameter,  30  ton  in  mass,  CuAl spherical  detector  operating  at
ultracryogenic  temperatures.   We  derived  the sensitivity  for  the
spherical detector in its initial  phase of development, when a single
transducer  is   used,  and  when  the  detector   operates  with  six
transducers  and becomes  fully omnidirectional.   The  sensitivity is
evaluated  when  the detector  operates  by  making  use of  available
technology and when it works at the quantum limit. We have shown that,
in order to  improve the strain sensitivity towards  the quantum limit
one should operate the detector  at temperature of about $T=20mK$ with
electrical and mechanical quality factor  as high as $5\cdot 10^7$ and
massive mechanical resonators.

Direction  anisotropies   in  the  detector   sensitivity  and  signal
bandwidth  are studied  for a  not-ideal detector  operating  with not
identical, partially tuned and  real-size resonators.  The models made
so  far always  consider  rather generic,  an point-like  transducers,
neglecting the  fact that those  sensors are in practice  rather large
and  spatially distributed  on a  significant fraction  of  the sphere
surface. We investigated here the validity of such an assumption.

Finally  we described  and numerically  verify a  complete calibration
procedure, which makes use  of techniques available for bar detectors.
Similar algorithms  can be used  on spherical detector  for diagnostic
purpose and to derive  the direction and polarization information from
the detected signal.

\begin{acknowledgments} The author would like to thank Massimo Bassan,
Florian Dubath,  Jean-Pierre Zendri and Alberto Lobo  for reading this
manuscript and for useful  discussions. The MiniGRAIL project has been
the  major  drive  for  writing   this  paper.   This  work  has  been
financially   supported  by   Integrated  Large   Infrastructures  for
Astroparticle Science (ILIAS) of  the Sixth Framework Programme of the
European Community.
\end{acknowledgments}

\bibliography{GOTTsphere}
\end{document}